\documentclass[%
twocolumn,english
reprint,
amsmath,amssymb,
aps,
pre
]{revtex4-2}
\usepackage[T1]{fontenc}
\usepackage[latin9,utf8]{inputenc}
\usepackage{graphicx}
\usepackage{esint}
\usepackage{xcolor}
\usepackage{bm}
\usepackage{hyperref}
\usepackage{natbib}

\makeatletter

\graphicspath{{images/}}
\usepackage{tikz}

\usepackage{siunitx}
\usepackage{booktabs}
\usepackage{array}
\newcolumntype{C}[1]{>{\centering\arraybackslash}p{#1}}

\newcommand{\spinup}{\Large{\textcolor{blue}{$\bm{\uparrow}$}}\normalsize}
\newcommand{\spindown}{\Large{\textcolor{red}{$\bm{\downarrow}$}}\normalsize}
\newcommand{\tikzcircle}[2][black,fill=gray]{\tikz\draw[#1,radius=#2] (0,0) circle ;}%
\newcommand{\psup}{\tikzcircle{4pt}}
\newcommand{\psdown}{\tikzcircle[fill=white]{4pt}}

\begin{document} 

\title{
Pseudotransitions in a dilute Ising chain
}

\author{Darya Yasinskaya}
\email{daria.iasinskaia@urfu.ru}
\affiliation{Department of Theoretical and Mathematical Physics, Institute of Natural Sciences and Mathematics, Ural Federal University, 19 Mira street, 620002 Ekaterinburg, Russia}
\author{Yury Panov}
\affiliation{Department of Theoretical and Mathematical Physics, Institute of Natural Sciences and Mathematics, Ural Federal University, 19 Mira street, 620002 Ekaterinburg, Russia}

\begin{abstract}
This study provides a comprehensive analysis of the ground state and thermodynamic properties of a spin-pseudospin chain representing a model of a one-dimensional dilute magnet with two types of nonmagnetic charged impurities. 
For this purpose, a method utilizing the transfer-matrix properties is employed. 
Despite the wide variety of intriguing frustrated phase states, we show that the model showcases pseudotransitions solely between simple charge and magnetic quasiorders. 
These pseudotransitions are characterized by distinct features in the thermodynamic and magnetic quantities, resembling first- and second-order phase transitions. In addition to pseudotransitions for the ``pure'' system, similar to those observed in other one-dimensional spin models, this study also reveals the presence of ``second-order'' pseudotransitions for the dilute case.
We show that the nature of these discovered pseudotransitions is associated with the phase separation in the chain into regions of (anti)ferromagnetic and charge-ordered phases. 
The ability to compare the results of an exact transfer-matrix calculation with a simple phenomenological description within the framework of Maxwell construction contributes to a deeper understanding of both the physical mechanisms underlying this phenomenon and the analytical methods used.
\end{abstract}


\maketitle

\section{Introduction}

The statistical solvability of one-dimensional models makes it possible to trace in detail the sources of occurrence of certain features of thermodynamic properties. 
This capability is particularly useful in assessing the applicability of various phenomenological constructions for more complex systems.

Despite the well-known rigorous theorems of the absence of traditional phase transitions in one-dimensional models with short-range interactions~\cite{vanHove1950Physica,Cuesta2004JSP}, pseudotransitions have been found in a wide class of (quasi)-one-dimensional systems. They manifest themselves as sharp features of specific heat, susceptibility, and correlation length at the finite temperature. This behavior resembles second-order phase transitions; however, for one-dimensional models, all functions remain continuous and have a finite, albeit very large, maximum value. The experimental detection of pseudotransitions is still under discussion. The search for pseudotransitions in real physical systems is compounded by the challenge that these phenomena exist only within a relatively narrow range of parameters. Although it is theoretically possible to design artificial materials with specified properties, such as through the use of qubits~\cite{OHare2007Aug}, this has yet to be successfully implemented. Nevertheless, the unique features of these phenomena could find a variety of applications~\cite{Yin2024PRR}.
In addition, pseudotransitions offer hope of approaching forbidden phase transitions at finite temperatures~\cite{Yin2024PRR,Yin2024PRB}. 

The existence of a pseudotransition obeys the rule formulated by Rojas~\cite{Rojas2020BJP,*Rojas2020APPA}: if the residual entropy is continuous at least from the one-sided limit at the boundary between the ground state phases, a pseudotransition at finite temperature will be observed near this boundary. At the microscopic level, this means~\cite{Panov2021PRE} that the ground state phases with the same energy at the boundary do not form a mixture with higher entropy than those of the neighboring phases. This is a rather rare situation that imposes significant constraints on the nature of the neighboring phases and makes pseudotransition a rare phenomenon. The high-temperature state in pseudotransitions is characterized by high entropy, but does not simply exhibit disorder. Unlike conventional phase transitions, this state is a frustrated quasiphase that does not mix with the low-temperature low-entropy state at the microscopic level.

Therefore, the presence of frustration is a crucial condition for the occurrence of pseudotransitions in the system. Previous studies of pseudotransitions have focused on exploring complex frustrated (quasi)-one-dimensional systems, including decorated chains~\cite{Rojas2019PRE,Krokhmalskii2021PhysicaA} such as diamondlike chains~\cite{Hovhannisyan2016JPCM,Carvalho2018JMMM,Carvalho2019AP}, ladders~\cite{Rojas2016SSC,Hutak2021PLA}, tubes~\cite{Strecka2016JMMM}, double-tetrahedral chains~\cite{galisova2015vigorous,Rojas2020JPCM}, and hexagonal nanowires~\cite{Pimenta2022JMMM}. 
Additionally, pseudotransitions have been observed in anisotropic Potts models with $q$ states~\cite{Panov2021PRE,Panov2023PRE}.
In general, the study of these models addresses the theoretical framework necessary for understanding a rapidly growing class of materials known as single-chain magnets~\cite{Coulon2006,Zhang2013}.

Actually, the source of frustrations in spin chains can be impurities. As a source of nonmagnetic impurities, in addition to chemical substitution, it is important to mention the photoinduced change of the magnetic state of ions in single-chain magnets~\cite{Liu2013}, as well as the charge disproportionation reaction that is characteristic of a wide class of magnets~\cite{Moskvin2023}.
However, it has not been previously observed that pseudotransitions can occur in dilute chains with impurities. The absence of pseudotransitions in the dilute Ising chain was earlier proved~\cite{Panov2022PRE}.
We assume that the lack of pseudotransitions in this dilute system is explained by the absence of frustrated phases with the necessary properties. In this work, we investigate an Ising spin chain that is diluted with nonmagnetic charged impurities of two types. These impurities are mobile and subject to density-density interactions. A two-dimensional version of this model was considered in~\cite{Panov2017JLTP,Panov2017JETPlett,Panov2019JMMM,Panov2019JSNM,Yasinskaya2020PSS,Yasinskaya2020APPA,Yasinskaya2021PSS,Yasinskaya2022IEEE}, as an atomic limit of a more general model for cuprates~\cite{Moskvin2011PRB,Moskvin2015JETP,Panov2019PMM,Moskvin2020PSS,Moskvin2021CM,Moskvin2022JMMM,Moskvin2022JPCS}. 
Despite the simplified nature of the spin-spin interaction, this model was used to describe the competition of charge and magnetic ordering in the normal phase of cuprates.

The valence multiplet of the many-electron states of the $[$CuO$_{4}]^{7-,6-,5-}$ centers, or nominally Cu$^{1+,2+,3+}$, corresponds to the three components of the pseudospin triplet.
The spin singlets Cu$^{1+}$ and Cu$^{3+}$ play the role of nonmagnetic centers having different charges and subject to density-density correlations, while the magnetic spin-$1/2$ centers of Cu$^{2+}$ are bound by antiferromagnetic superexchange interaction. 
Note that a similar model has been used for the one-dimensional cuprate chains~\cite{Strecka2024EPJB}, which can serve as a prototype for one-dimensional cuprates such as SrCuO$_{2}$~\cite{Bounoua2017}.

The peculiarity of our model is accounting for the condition of constant charge concentration in the system. This not only requires the use of specialized techniques to calculate thermodynamic properties, but also results in novel features of pseudotransitions. Similar to systems with a conserved order parameter~\cite{Yang1952PR,Newman}, the lattice hardcore bosons or the extended Hubbard models~\cite{Batrouni2000PRL,Kapcia2012JPCM,Kapcia2013JPCM,Kapcia2013JSNM,Kapcia2015JSNM,Kapcia2015PhysicaA,Kapcia2017PRE}, our model exhibits various forms of phase separation.
Analytical treatment of phase separation in real systems is a complex task, and often involves the use of phenomenological assumptions, such as Maxwell construction~\cite{Plischke2006}. 
The study of our one-dimensional model allows us to assess the validity of using and the limits of applicability of such constructions. 

The paper is structured as follows. 
Section~II presents the spin-pseudospin model of a dilute Ising chain and specifics of the calculation methods. 
Properties of the ground state phases are discussed in Section~III, 
and the most interesting results about conventional pseudo-transitions in a ``pure'' chain and the previously unobserved type of pseudotransitions in a dilute chain, along with their interconnections, are presented in Section~IV.
Section~V summarizes our key findings and draws conclusions. 
Some additional details are given in the Appendix.

\section{Model and methods}

The Hamiltonian of the one-dimensional Ising model with charged impurities has the following form:
\begin{multline}
	\mathcal{H} = \sum_{i=1}^N \left\{ \Delta  S_{z,i}^2 + V S_{z,i} S_{z,i+1}\right.\\
	 + \left.J P_{0,i} s_{z,i} s_{z,i+1} P_{0,j} - h P_{0,i} s_{z,i} - \mu S_{z,i} \right\}.
\label{H}
\end{multline}
The summation is carried out over $N$ sites of a one-dimensional chain. 
Here in terms of the pseudospin $S=1$ triplet we express the charge states. Nonmagnetic impurities might have positive or negative charges with $S_z = \pm 1$, respectively. The projection operator $P_{0,i} = 1 - S^2_{z,i}$ distinguishes the magnetic state with $S_z = 0$, which is described by the Ising spin $s = 1/2$. $\Delta$ corresponds to the on-site density-density correlation, which takes the form of single-ion anisotropy for pseudospins; $V > 0$ denotes the inter-site density-density interaction coupling for impurities in the form of the Ising exchange coupling for pseudospins; $J$ represents the Ising exchange coupling for spins; $h$ is the external magnetic field; $\mu$ is the chemical potential for impurities. Here, for simplicity, we will assume magnetic field to be zero. Turning to the concentration of the impurities, one can write the condition of conservation of charge density:
\begin{equation}
	n = \langle S_z \rangle = const.
\end{equation}
A more detailed discussion of the form of density-density interaction within the framework of the pseudospin formalism was given in~\cite{Yasinskaya2021PSS}. 
Thus, each node of the chain can be either in one of the charged states (with spinless pseudospin states $S_z = \pm 1$, representing positively and negatively charged impurities, respectively) or in one of the magnetic states (with spin states $s_z = \pm 1/ 2$, corresponding to the pseudospin projection $S_z = 0$). 
We will assume that nonmagnetic impurities are mobile, representing an annealed system.

The expressions for the grand potential per node for different ground state phases are as follows:
\begin{eqnarray}
	&&\omega_{\text{I}}^{\pm} = \Delta + V \pm \mu, \quad
	\omega_{\text{CO}} = \Delta - V, \quad 
	\omega_{\text{PM}}^{\pm} = \frac{\Delta \pm \mu}{2}, 
	\nonumber\\
	&&\omega_{\text{FM}} = J, \quad 
	\omega_{\text{AFM}} = -J .
	\label{gpot}
\end{eqnarray}

Impurity (I), ferromagnetic (FM), antiferromagnetic (AFM), staggered charge order (CO) and paramagnetic (PM) phases correspond to the following configurations of nearest neighbors states: I$^{\pm}$ $\rightarrow$ $(\pm 1,\pm 1)$, FM $\rightarrow$ $\left(\frac{1}{2}, \frac{1}{2}\right)$, AFM $\rightarrow$ $\left(\frac{1}{2}, -\frac{1}{2}\right)$, CO $\rightarrow$ $(1,-1)$, PM$^{\pm}$ $\rightarrow$ $\left(\pm 1, \frac{1}{2}\right)$. 
These ``pure'' phases are characterized by the following charge densities: $n_{\text{I}^{\pm}} = \pm 1$, $n_{\text{FM}} = n_{\text{AFM}} = n_{\text{CO}} = 0$, $n_{\text{PM}^{\pm}} = \pm \frac{1}{2}$.

Transfer-matrix of a one-dimensional system with Hamiltonian~\eqref{H} in a local basis of states $\Phi = \left\{ \left|S_z, s_z\right\rangle \right\} $ $= \left\{\left|+1,0\right\rangle, \left|-1,0\right\rangle, \left|0,+\frac{1}{2}\right\rangle, \left|0,-\frac{1}{2}\right\rangle \right\} $ $\equiv \left\{+1, -1, +\frac{1}{2}, -\frac{1}{2} \right\}$ has the structure
\begin{equation}
	\hat{T} = \begin{pmatrix}
		e^{-\beta \omega^{-}_{\text{I}} } & e^{-\beta \omega_{\text{CO}}} & e^{-\beta \omega^{-}_{\text{PM}} } & e^{-\beta \omega^{-}_{\text{PM}} } \\
		e^{-\beta \omega_{\text{CO}}}  & e^{-\beta \omega^{+}_{\text{I}} } & e^{-\beta \omega^{+}_{\text{PM}} } & e^{-\beta \omega^{+}_{\text{PM}} } \\
		e^{-\beta \omega^{-}_{\text{PM}} } & e^{-\beta \omega^{+}_{\text{PM}} } & e^{-\beta \omega_{\text{FM}} } & e^{-\beta \omega_{\text{AFM}} } \\
		e^{-\beta \omega^{-}_{\text{PM}} } & e^{-\beta \omega^{+}_{\text{PM}} } & e^{-\beta \omega_{\text{AFM}} } & e^{-\beta \omega_{\text{FM}} } \\
	\end{pmatrix},
	\label{TM}
\end{equation}
where $\beta = 1/\left(k_B T\right)$, $k_B = 1$.

Using the technique of the transfer-matrix, the partition function can be calculated from the transfer-matrix eigenvalues $\{\lambda_i, \, i=1 ... 4\}$:
\begin{equation}
	Z = \text{Tr} \left(e^{-\beta \mathcal{H}} \right) = \text{Tr} \left(\hat{T}^N \right) = \sum\limits_{i=1}^4 \lambda_i^N . 
\end{equation}
The detailed expressions for these eigenvalues are provided in the Appendix. 
According to the Perron-Frobenius theorem~\cite{Frobenius1912}, there is only one maximum eigenvalue, $\lambda_1$, and in the thermodynamic limit, the grand potential can be expressed as follows:
\begin{equation}
	\omega = \frac{\Omega}{N} = -T \ln \lambda_1 . 
\end{equation}

The standard transfer-matrix method further assumes that all thermodynamic quantities, including entropy, specific heat, magnetization, magnetic susceptibility, etc., as well as local correlators and pair correlation functions, are derived through the eigenvalue $\lambda_1$.

However, the expression for $\lambda_1$ given in Appendix is quite cumbersome. 
To calculate thermodynamic properties, it is necessary to manipulate even more intricate expressions involving various derivatives of $\lambda_1$. 
Consequently, this becomes more challenging and time-consuming, especially at low temperatures. This problem exacerbates when the need arises to proceed to a fixed charge density of impurities $n \neq 0$. 
In this case, it is imperative to numerically solve the equation $n(\mu) = n$ for each temperature and parameter set to determine the appropriate chemical potential $\mu$. 
In light of this challenge, we propose to work with expectation values of various thermodynamic functions, which can be directly obtained using matrices built on eigenvectors of a transfer matrix.

Consider an operator satisfying the relation
\begin{equation}
	A = \frac{\partial \left( -\beta \mathcal{H} \right)}{\partial \alpha},
	\label{relation}
\end{equation}
which has the matrix form $\hat{A}$ in the eigenbasis of the transfer-matrix $\hat{T}$.

In that case, the expectation value of this operator can be expressed as follows
\begin{equation}
	\langle A \rangle = \frac{\partial \ln Z}{\partial \alpha} = \frac{1}{Z} \frac{\partial \text{Tr} \left(\hat{T}^N\right)}{\partial \alpha} = \frac{N}{Z} \text{Tr} \left( \frac{\partial \hat{T}}{\partial \alpha}\hat{T}^{N-1} \right),
\end{equation}
where cyclic permutation under the trace is used. Transforming into a basis of eigenvectors, one can calculate the expectation value from the transfer-matrix eigenvalues and the elements of matrix $\frac{\partial \hat{T}}{\partial \alpha}$ in the eigenbasis:
\begin{equation}
	\langle A \rangle = \frac{N \sum\limits_i \left(\frac{\partial \hat{T}}{\partial \alpha}\right)_{ii} \lambda_i^{N-1}}{\sum\limits_i  \lambda_i^{N}}.
\end{equation}
Then, in the thermodynamic limit we obtain
\begin{equation}
	\frac{\langle A \rangle }{N} \xrightarrow[N \to \infty]{} \frac{1}{\lambda_1} \left(\frac{\partial \hat{T}}{\partial \alpha}\right)_{11} . 
\end{equation}

For two commuting operators $A$ and $B$, where $B~=~\frac{\partial (-\beta \mathcal{H})}{\partial \gamma}$ and has a matrix $\hat{B}$ in the eigenbasis, the expectation value of their product is also expressed in terms of a transfer-matrix:
\begin{eqnarray}
	\langle A B \rangle &=& \frac{\text{Tr} \left( \hat{A} \hat{B} e^{-\beta \mathcal{H}} \right)}{\text{Tr} \left( e^{-\beta \mathcal{H}} \right)} \label{ABprod} \\* \nonumber
	&=&\frac{\text{Tr} \left(\frac{\partial^2 \hat{T}}{\partial \alpha \partial \gamma} \hat{T}^{N-1}\right) + \sum\limits_{k=0}^{N-2}\text{Tr} \left( \frac{\partial \hat{T}}{\partial \alpha} \hat{T}^k \frac{\partial \hat{T}}{\partial \gamma} \hat{T}^{N-2-k} \right)}{\text{Tr} \left(\hat{T}^N \right)} . 
\end{eqnarray}
In the eigenbasis the sum of traces in~\eqref{ABprod} collapses into the sum of a geometric series:
\begin{widetext}
\[
\begin{split}
	&\sum\limits_{k=0}^{N-2}\text{Tr} \left( \frac{\partial \hat{T}}{\partial \alpha} \hat{T}^k \frac{\partial \hat{T}}{\partial \gamma} \hat{T}^{N-2-k} \right) = \sum\limits_{k=0}^{N-2} \sum\limits_{ij} \left(\frac{\partial \hat{T}}{\partial \alpha}\right)_{ij} \lambda_i^k \left(\frac{\partial \hat{T}}{\partial \gamma}\right)_{ji} \lambda_j^{N-2-k} \\*
	&= (N-1) \sum\limits_i \left(\frac{\partial \hat{T}}{\partial \alpha}\right)_{ii} \left(\frac{\partial \hat{T}}{\partial \gamma}\right)_{ii} \lambda_{i}^{N-2} + \sum\limits_{i \neq j} \left(\frac{\partial \hat{T}}{\partial \alpha}\right)_{ij} \left(\frac{\partial \hat{T}}{\partial \gamma}\right)_{ji} \frac{\lambda_i^{N-1} - \lambda_j^{N-1}}{\lambda_i - \lambda_j} . 
\end{split}
\]
\end{widetext}
In the thermodynamic limit, the leading elements will correspond to the largest eigenvalue:
\begin{multline}
	\frac{\langle AB \rangle}{N} 
	\xrightarrow[N \to \infty]{} \frac{1}{\lambda_1} \left(\frac{\partial^2 \hat{T}}{\partial \alpha \partial \gamma}\right)_{11} 
	+ \frac{N-1}{\lambda_1^2} \left(\frac{\partial \hat{T}}{\partial \alpha}\right)_{11} \left(\frac{\partial \hat{T}}{\partial \gamma}\right)_{11} \\
	+ 	 \frac{2}{\lambda_1^2} \sum\limits_{j \neq 1} \frac{\left(\frac{\partial \hat{T}}{\partial \alpha}\right)_{1j} \left(\frac{\partial \hat{T}}{\partial \gamma}\right)_{j1}}{1 - \frac{\lambda_j}{\lambda_1} } . 
\end{multline}
Finally, the covariance of two commuting operators $A$ and $B$ is as follows:
\begin{multline}
	\frac{\langle AB \rangle - \langle A \rangle \langle B \rangle}{N} \xrightarrow[N \to \infty]{}  \frac{1}{\lambda_1} \left(\frac{\partial^2 \hat{T}}{\partial \alpha \partial \gamma}\right)_{11} \\
	 -\frac{1}{\lambda_1^2} \left(\frac{\partial \hat{T}}{\partial \alpha}\right)_{11} \left(\frac{\partial \hat{T}}{\partial \gamma}\right)_{11} + \frac{2}{\lambda_1^2} \sum\limits_{j \neq 1} \frac{\left(\frac{\partial \hat{T}}{\partial \alpha}\right)_{1j} \left(\frac{\partial \hat{T}}{\partial \gamma}\right)_{j1}}{1 - \frac{\lambda_j}{\lambda_1} } . 
\end{multline}
Diagonalizing the transfer matrix with a specified level of accuracy, computing the derivative matrices in the eigenbasis, and manipulating the corresponding matrix elements is a more efficient approach compared to handling complex expressions for $\lambda_1$ and its derivatives.

Accordingly, entropy, magnetization and charge density can be expressed through the expectation values of certain functions such as the Hamiltonian~\eqref{H}, total spin and pseudospin projections, respectively:
\begin{multline}
	\mathcal{S} = \frac{1}{NT} \left[ - \Omega + \langle \mathcal{H} \rangle\right] = - \left(\frac{\partial \omega}{\partial T} \right)_{h,\mu} \\= \frac{1}{N} \left[ \ln Z + \frac{T}{Z} \text{Tr}\left(\frac{\partial \hat{T}}{\partial T} \hat{T}^{N-1} \right) \right] ,
\end{multline}
\begin{equation}
	m = \langle s_z \rangle = -\left(\frac{\partial \omega}{\partial h} \right)_{T,\mu} = \frac{T}{N Z} \text{Tr}\left(\frac{\partial \hat{T}}{\partial h} \hat{T}^{N-1} \right),
	\label{eq:m}
\end{equation}
\begin{equation}
	n =  \langle S_z \rangle = -\left(\frac{\partial \omega}{\partial \mu} \right)_{T,h} =  \frac{T}{N Z} \text{Tr}\left(\frac{\partial \hat{T}}{\partial \mu} \hat{T}^{N-1} \right) .
\end{equation}
These functions, in turn, satisfy the relation~\eqref{relation} as derivatives with respect to temperature, magnetic field and chemical potential. 
By fixing the charge density $n$, we can numerically determine the chemical potential $\mu$ as a solution of the self-consistent equation 
\begin{equation}
	n(\mu) = n
\end{equation}
and calculate all other thermodynamic quantities.
Subsequently, the specific heat and magnetic susceptibility in the grand canonical formalism for fixed values of $h$ and $n$ are expressed through the covariances as follows:
\begin{multline}
	C_{h,n} = T \left[ \left(\frac{\partial \mathcal{S}}{\partial T}\right)_{h,\mu} - \frac{\left(\frac{\partial n }{\partial T}\right)_{h,\mu}^2}{\left(\frac{\partial n }{\partial \mu}\right)_{T,h}} \right] \\
	=\frac{1}{N^2 T^2} \left[ \langle \mathcal{H}^2 \rangle - \langle \mathcal{H} \rangle^2 - \frac{\left( \langle \mathcal{H} S_z \rangle - \langle \mathcal{H} \rangle \langle S_z \rangle \right)^2}{\langle S_z^2 \rangle - \langle S_z \rangle^2} \right] . 
\end{multline}
\begin{multline}
	\chi_{T,n} = \left(\frac{\partial m}{\partial h}\right)_{T,\mu} - \frac{\left(\frac{\partial n }{\partial h}\right)_{T,\mu}^2}{\left(\frac{\partial n }{\partial \mu}\right)_{T,h}} \\
	= \frac{1}{T} \left[ \langle s_z^2 \rangle - \langle s_z \rangle^2 - \frac{\left( \langle s_z S_z \rangle - \langle s_z \rangle \langle S_z \rangle \right)^2}{\langle S_z^2 \rangle - \langle S_z \rangle^2} \right] . 
\end{multline}

We consider the following order parameters for this system:
\begin{equation}
	\mathcal{O} = \begin{cases}
		m_\text{FM} = m,\\
		m_\text{AFM} = m_1 - m_2 ,\\
		M_\text{CO} = M_1 - M_2,\\
	\end{cases}		
	\label{eq:op}
\end{equation}
where $m_\text{FM}$ is magnetization for ferromagnetic order determined by Eq.~\eqref{eq:m}; 
$m_{\lambda}~=~\frac{1}{N} \left| \sum\limits_{i \in \lambda} s_{iz} \right| $ is the magnetization of a sublattice $\lambda = 1,2$, so that $m_\text{AFM}$ is an antiferromagnetic vector; 
$M_{\lambda}~=~\frac{1}{N} \left| \sum\limits_{i \in \lambda} S_{iz} \right| $ is the summary charge of a sublattice $\lambda$ (pseudo-magnetization), so that $M_\text{CO}$ is the checkerboard charge order parameter.

Furthermore, employing the method of calculating expectation values from a transfer matrix enables the direct computation of pair correlation functions. Consequently, spatial correlators for two operators $A$ and $B$ will take on a specific form in the thermodynamic limit:
\begin{multline}
	\langle A_i B_{i+d} \rangle = \frac{\text{Tr} \left( \hat{T}^i \hat{A} \hat{T}^d \hat{B} \hat{T}^{N-d} \right) }{\text{Tr} \left(\hat{T}^N \right)}\\  \xrightarrow[N \to \infty]{} \sum_{k=1}^4 (\hat{A})_{1k} (\hat{B})_{k1} \left(\frac{\lambda_k}{\lambda_1} \right)^d,
\end{multline}
where $d = \vert j-i \vert$ is the distance between nodes $i$ and $j$. 
Therefore, only the elements of the first columns and rows of the operator matrices $\hat{A}$ and $\hat{B}$ in the eigenbasis contribute to the expression.

\section{Ground state properties: diagrams, entropy and phase boundaries}

\begin{figure*}
	\includegraphics[width=\linewidth]{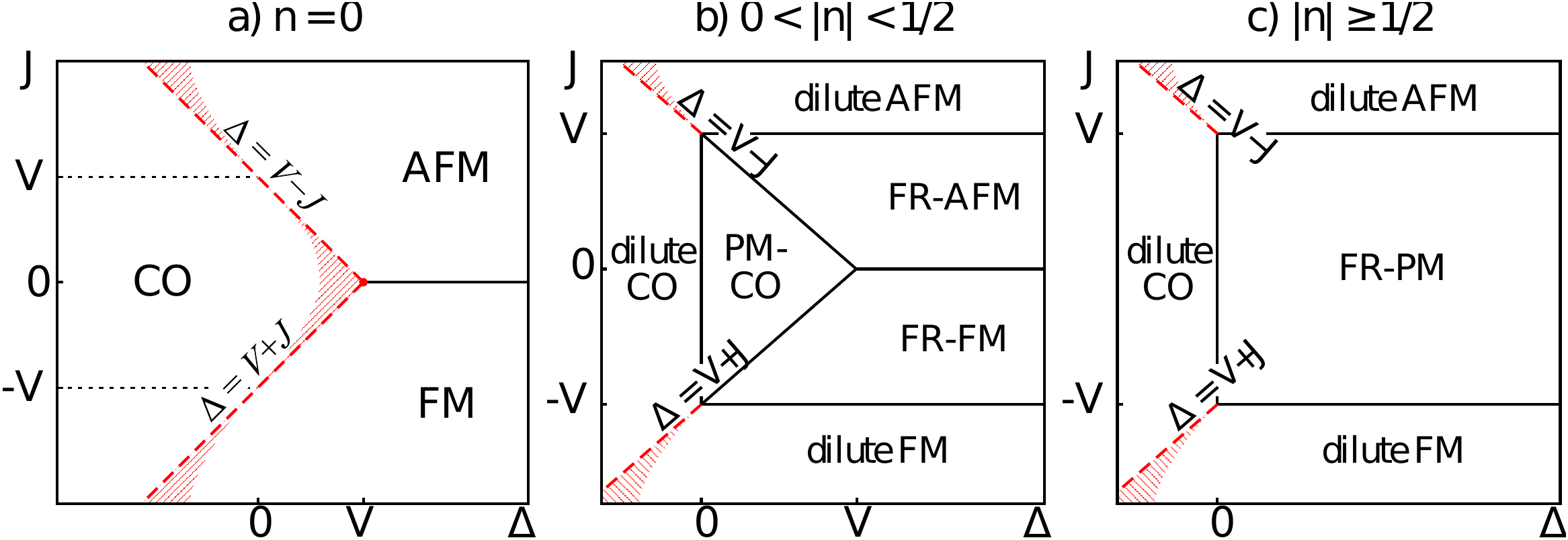}
	\caption{Ground state phase diagrams in the $\Delta - J$ plane depend on charge density $n$. 
	The red shading schematically indicates the regions where pseudotransitions are supposedly observed. Accordingly, the phase boundaries indicated by the red dashed lines satisfy the Rojas criterion~\cite{Rojas2020BJP}.}
	\label{GSPD}
\end{figure*}

\begin{figure}
	\centering
	\includegraphics[width=0.75\linewidth]{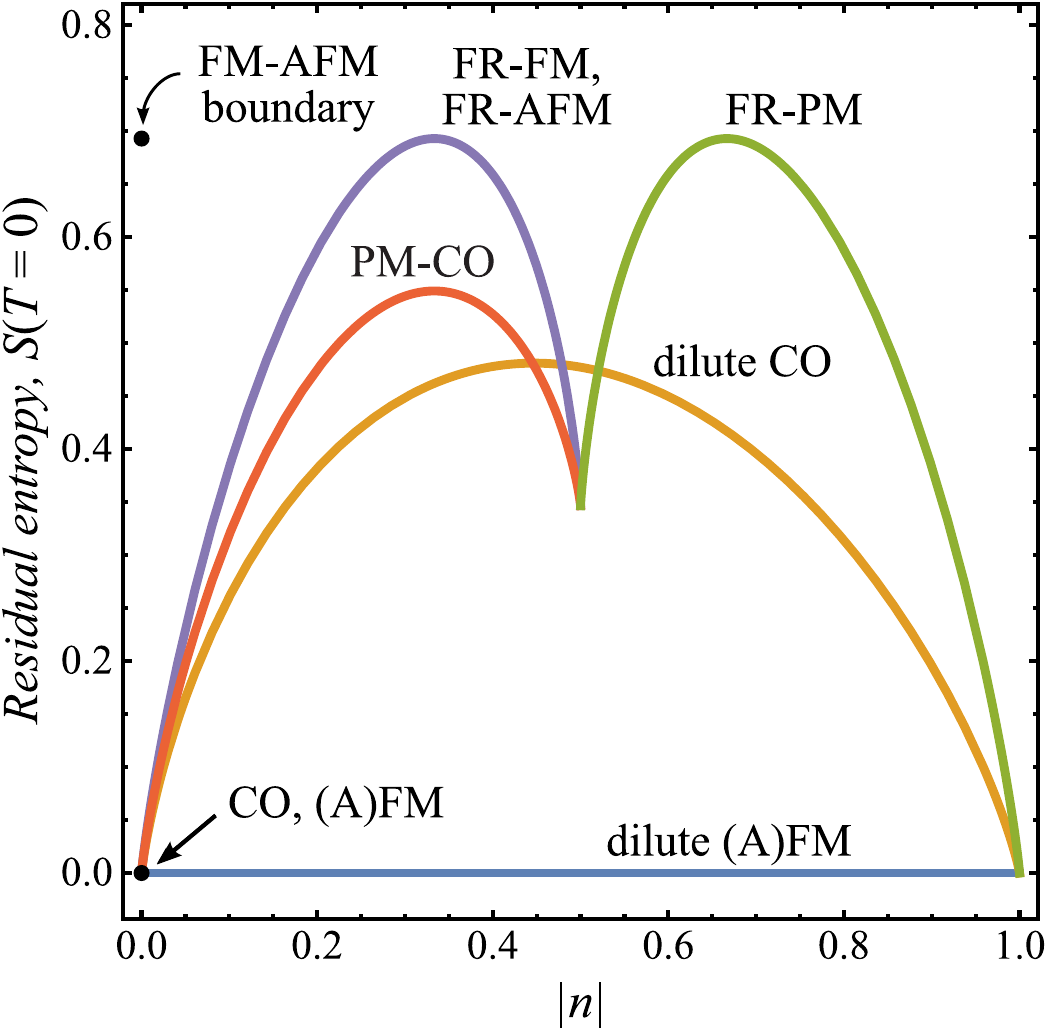}
	\caption{Dependency of the residual entropy of various phases on the charge density of nonmagnetic impurities $n$. 
		The (anti)ferromagnetic phase is the only phase with zero entropy and is nonfrustrated.}
	\label{s(n)}
\end{figure}

Let us now proceed to the discussion of a wide variety of ground state phases in a dilute Ising magnet. 
All the ground states are presented in Table~\ref{phtab} with their energies and corresponding chain state illustrations. 
These ground states collectively form the ground state phase diagrams (GSPD) shown in Fig.~\ref{GSPD} in the $(\Delta,J)$ plane for three cases of $n$: (a) ``pure'', $n=0$; (b) weak dilution, $0 < \left| n \right| < 1/2$; and (c) strong dilution, $ \left| n \right| \geq 1/2$. 
Some of these phases exhibit frustration due to the presence of numerous permutations of impurities that do not change the system's energy. This is manifested as nonzero residual entropy, which depends on charge density $n$, as is shown in Fig.~\ref{s(n)}. 
Now examine the properties of these ground state phases in more detail.

\begin{table*}
	\renewcommand{\tabcolsep}{0.5cm}
	\centering
	\caption{Ground state phases of a dilute Ising chain, their energy, and corresponding chain state. 
	For simplicity, all phases are depicted for $n \geq 0$ in the presence of a magnetic field $h>0$, which dictates the preferred direction for spin orientation. 
	Gray dots (\psup) and white dots (\psdown) denote the two charge states of pseudospin, $S_z = \pm 1$. 
	Blue (\spinup) and red (\spindown) arrows denote the two magnetic states of spin, $s_z = \pm 1/2$, which correspond to $S_z = 0$}
	\label{phtab}
	\begin{tabular}{ |c |  c | c | c |}
		\hline
		N. & Phase & Energy & Typical Configurations \\
		\hline
		1 & (dilute) FM & $( J - \left| h \right| )(1 - \left| n \right| ) + \left| n \right| (\Delta + V)$ & \spinup \spinup  \spinup \spinup \spinup \psup \psup \psup \psup \spinup \spinup \spinup 
		\\
		\hline
		2 & (dilute) AFM & $ - J(1 -  \left| n \right|) +  \left| n \right| (\Delta + V)$ & \spindown \spinup \spindown \spinup \spindown \psup \psup \psup \psup \spinup \spindown \spinup 
		\\
		\hline
		3 & (dilute) CO & $\Delta - V(1 - 2 \left| n \right|)$ & \psup \psup \psdown \psup \psdown \psup \psdown \psup \psup \psdown \psup \psdown
		\\
		\hline
		4 & PM-CO & $- V (1 - 2 \left| n \right|  ) + \Delta (1 -\left| n \right|) - \left| n \right| \vert h \vert$ &  \psup \psdown \psup \spinup \psup \psdown \psup \psdown \psup \spinup \psup \spinup 
		\\
		\hline
		5 & FR-AFM & $- J(1 - 2 \left| n \right|)+\left| n \right|(\Delta-\vert h \vert)$ & \spinup \spindown \spinup \psup \spinup \psup \spinup \spindown \spinup \spindown \spinup \psup \spinup 
		\\
		\hline
		6 & FR-FM &  $J(1-2 \left| n \right|) - \vert h \vert (1 - \left| n \right|) + \Delta \left| n \right|$ & \spinup \psup \spinup \spinup \spinup \spinup \psup \spinup \psup \spinup \spinup \spinup \psup
		\\
		\hline
		7 & FR-PM & $- V(1 - 2 \left| n \right|) - \vert h \vert (1 - \left| n \right|) + \Delta \left| n \right| $ &  \psup \spinup \psup \psup \psup \spinup \psup \spinup \psup \spinup \psup \psup
		\\
		\hline
	\end{tabular}
\end{table*}

In a ``pure'' spin chain, $n=0$, there are only three ground states, as illustrated in Fig.~\ref{GSPD}(a).
The ``checkerboard'' CO state exhibits a regular alternation of negatively and positively charged impurities. The energy per node in the CO phase remains independent of the exchange coupling parameter $J$. 
In the FM phase all spins are aligned in one direction. The AFM phase is a classical N\`eel state. These three phases have zero residual entropy: $\mathcal{S}_{\text{CO, FM, AFM}} (T=0) = 0$.

\begin{figure}
	\centering
	\includegraphics[width=0.75\linewidth]{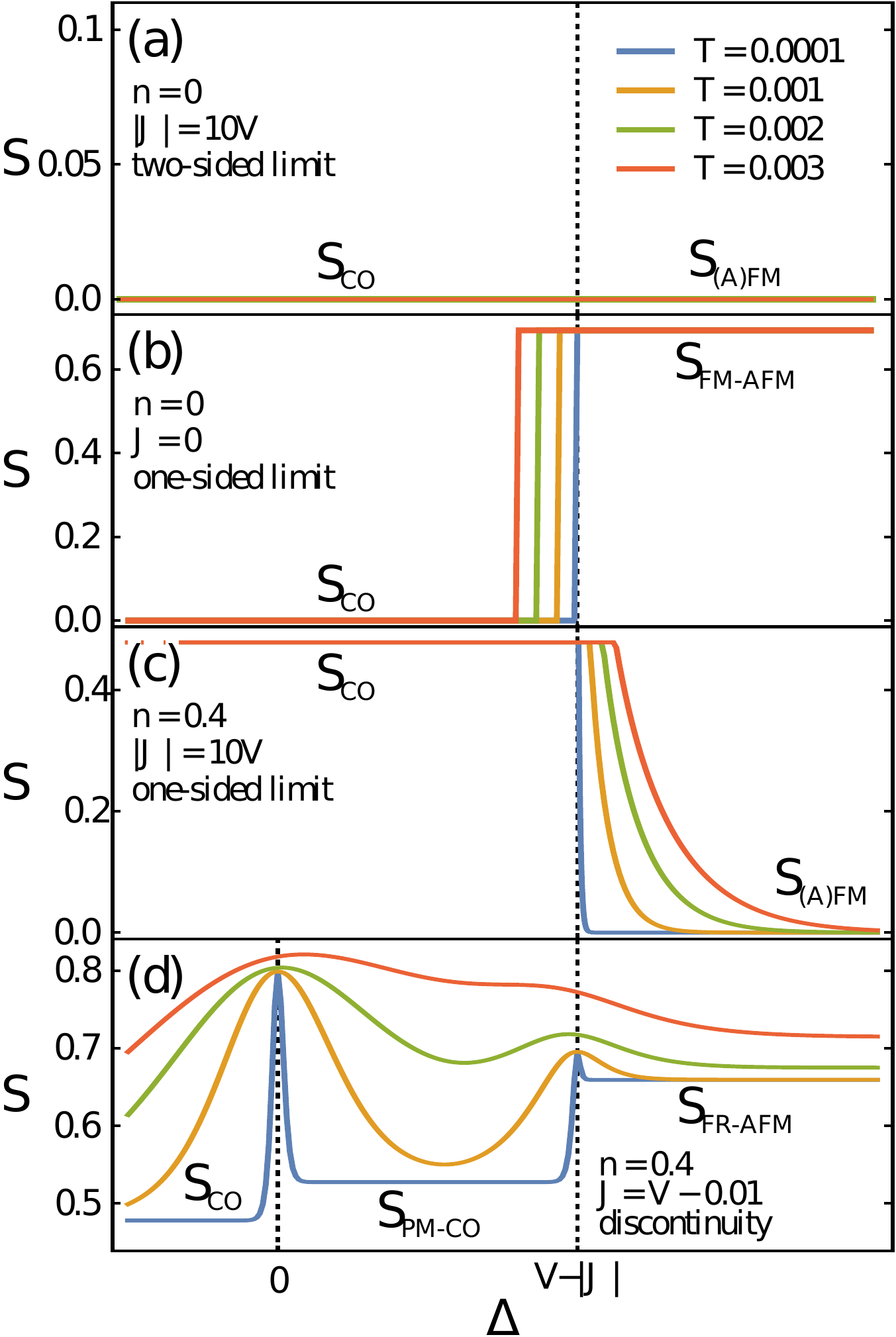}
	\caption{Dependencies of the entropy on $\Delta$ at low temperatures $T=10^{-4}$, $10^{-3}$, $2 \cdot 10^{-3}$, $3\cdot 10^{-3}$ near different phase boundaries. 
	Typical scenarios of continuous and discontinuous residual entropy on these boundaries are represented: 
	(a) continuous residual entropy equal zero from both CO and (A)FM sides, with $n=0$, $\vert J \vert = 10V$; 
	(b) continuous residual entropy at the triple point CO--FM--AFM from the one-sided limit, $n=0$, $J=0$; 
	(c) continuous residual entropy at the boundary CO--(A)FM from the one-sided limit, $n=0.4$, $\vert J \vert = 10 V$; 
	(d)  discontinuous residual entropy at the PM-CO--FR-AFM boundary, $n=0.4$, $\vert J \vert = V/2$. 
	Only cases (a)--(c) suggest pseudotransitions based on entropy criterion~\cite{Rojas2020BJP}}
	\label{s(D)}
\end{figure}

The boundaries between charge and magnetic phases are determined by the conditions $\Delta = V - \left|J\right| < V$ and depicted in Fig.~\ref{GSPD}(a) by the red dashed lines. 
It is important to note that CO and (A)FM phases coexist only if the density-density repulsion is counterbalanced by the spin exchange interaction, $V = \left| J \right|$. 
In that case, the system decomposes into magnetically and charge-ordered clusters, but the total residual entropy still remains equal to zero. 
In case of the strong spin exchange coupling, $V < \left| J \right|$, the state at the boundary is the CO, so that a two-sided limit for residual entropy can be observed, and pseudotransitions can be expected from the (A)FM side, according to the Rojas criterion~\cite{Rojas2020BJP}. 
The dependency of the entropy at $T=10^{-4}$ near this boundary is shown in Fig.~\ref{s(D)}(a). 

Conversely, in the weak exchange limit ($V > \left| J \right|$), a symmetrical scenario unfolds with a magnetic state forming at the CO--(A)FM boundary. 
A two-sided limit is also observed, but a pseudotransition is anticipated from the CO side. 
The regions exhibiting pseudotransitions are schematically highlighted by red shading on the GSPD in Fig.~\ref{GSPD}. 
Of particular interest is the triple point (CO--FM--AFM), denoted in red in Fig.~\ref{GSPD}(a).
Approaching this point along the line $J=0$ reveals a jump discontinuity in residual entropy, as depicted in Fig.~\ref{s(D)}(b). 
Consequently, a one-sided limit for residual entropy is observed. 
The state at the triple point, as well as at the FM--AFM boundary, is a frustrated mixture of FM and AFM clusters with residual entropy $\mathcal{S} = \ln 2$.

The above-discussed CO, (A)FM phases exist at any charge density $n$, but their diluted versions exhibit distinct properties. 
Charge impurities are able to coexist with CO; thus, impurities are randomly distributed throughout the system. 
This indicates the frustration and nonzero residual entropy of the dilute CO phase, which is dependent on $n$ given by expression
\begin{multline}
	 \mathcal{S}(n) = \frac{1+\left| n \right|}{2} \ln \frac{1+\left| n \right|}{2} \\
	- \frac{1-\left| n \right|}{2} \ln \frac{1-\left| n \right|}{2} - \left| n \right| \ln \left| n \right|,
	 \label{eq:S(n)}
\end{multline}
as depicted in Fig.~\ref{s(n)} by the orange line.

On the other hand, phase separation into magnetic domains and charged droplets occurs in the dilute magnetic phases FM and AFM. These charged droplets are macroscopic regions with a total volume $\left| n \right|$ and consist only of nodes occupied by impurities. 
The phenomenon of phase separation is observed when the spin exchange interaction is stronger than the charge repulsion, $\left| J \right| > V$. 
In such a case, the decrease in energy of spins interaction outweighs the energy loss due to the repulsion of like charges. The number of permutations of charged droplets in a chain that do not alter the ground state energy follows a power-law asymptotic behavior. This means that the residual entropy of dilute magnetic phases tends towards zero in the thermodynamic limit, $N \rightarrow \infty$. 
A detailed discussion of the phase separation in a two-dimensional dilute model was made in our earlier works using Monte Carlo computer simulation~\cite{Panov2017JETPlett}, as well as mean field calculation within the framework of Maxwell construction~\cite{Panov2019JMMM}.

The boundaries between the charge and magnetic phases are now only observed in the case of strong spin exchange, $\left|J\right| > V$, $\Delta = V - \left|J\right|$, as illustrated by the red dashed lines in Figs.~\ref{GSPD}(b,c). 
Nevertheless, the state at this boundary will result in a high-entropy dilute CO, leading to a one-sided limit for residual entropy, as again observed in Fig.~\ref{s(D)}(c). 
Pseudotransitions can also be anticipated in regions highlighted with red shading in Fig.~\ref{GSPD}.

If the spin exchange interaction is weaker than the charge repulsion,  $\left|J\right| < V$, nonmagnetic impurities do not aggregate into charged droplets but are randomly distributed throughout the system. As a result, the frustrated magnetic phases FR--(A)FM are established. These phases exist for $0 < \left| n \right| < \frac{1}{2}$ and exhibit nonzero residual entropy within this range, as depicted by the purple line for FR-FM and the red line for FR-AFM in Fig.~\ref{s(n)}.

The PM-CO and FR-AFM phases are, in some sense, symmetrical with respect to the substituting of spin states with pseudospin ones. As a result, the residual entropies of these two phases are identical. The PM-CO state is a dilute checkerboard charge order with paramagnetic centers represented by single spins. This phase is observed only at low charge densities, $0 < \left| n \right| < 1/2$. The energy of this phase does not depend on the exchange interaction constant $J$.

Upon reaching $\left| n \right| = 1/2$, the frustrated paramagnetic phase FR-PM, emerges, replacing both the FR-(A)FM and PM-CO phases. This phase represents the final frustrated phase observed under the conditions examined in this study. In the FR-PM phase, single spins are separated by clusters of nonmagnetic impurities with the same charge. As a result, the system displays a paramagnetic response and retains a nonzero residual entropy. The symmetry with the FR-FM phase is clearly evident when substituting pseudospin states with spin states. The residual entropy is symmetrical to $S_{\text{FR-FM}}$ about the point $ \left| n \right| = 1/2$, as shown by the green line in Fig.~\ref{s(n)}.

While there is a diverse range of frustrated phase states, only the most basic ones, such as the CO and (A)FM phases, meet the necessary conditions for pseudotransitions to take place. The boundaries between all other phases exhibit higher residual entropy than that of the individual phases, resulting in discontinuities. This is illustrated in Fig.~\ref{s(D)}(d), showcasing an example of a removable discontinuity of the residual entropy at the boundary between the PM-CO and FR-AFM phases.

Moving forward, we will focus mostly on the boundary between the CO and (A)FM phases. To facilitate discussion, we will introduce the offset value from this ground state phase boundary:
\begin{equation}
	\delta \varepsilon_{0} = \varepsilon_{\text{CO}}^0 (0) - \varepsilon_{\text{(A)FM}}^0(0)
	= \Delta - V + \left| J \right|,
\end{equation}
where index 0 indicated $T=0$, and the charge density value is set to be 0.

Let us briefly discuss the prospective impact of the magnetic field on the ground state properties. First of all, a strong magnetic field facilitates the emergence of a new ground state phase. In this phase, spins are aligned with the magnetic field, alternating with nonmagnetic impurities. Consequently, impurities and spins do not have neighbors of the same type. This phase is inherently frustrated, even in a ``pure'' state with $n=0$, and is realized only for weak dilution case $0 \leq \left| n \right| < 1/2$. 
This phase borders the CO phase, effectively excluding the triple point CO-FM-AFM for $n=0$ and the region of pseudotransitions with $V > \left| J \right|$; therefore, pseudotransitions will be observed only when $\left| J \right| > V$. 
Second, the introduction of a magnetic field will result in the displacement of certain phase boundaries, which will affect the positioning and pseudocritical temperatures of pseudotransitions. Significantly, the magnetic field nontrivially affects the structure of the spin chain. It can change the type of a periodic Markov chain corresponding to the magnetic ground state, and potentially lead to the emergence of sublattice long-range order~\cite{Panov2023PRE,Yasinskaya2024PSS}. 
However, it should be emphasized that the presence of a magnetic field is not imperative for the main focus of this study.

\section{Thermodynamics of a dilute Ising model}

\subsection{``First-order'' pseudotransitions}

Next, let us delve into a comprehensive analysis of the temperature properties of the ``pure'' system at $n = 0$ and focus on the emergence of pseudotransitions in this scenario. 
As outlined in the previous section, we anticipate pseudotransitions to occur near the boundary between the charge and magnetic phases based on the entropy criterion, with the exception of the critical point $V=\left|J\right|$. 
The range of parameters where pseudotransitions are observed depends on the proximity to this critical point, with range becoming significantly narrower at $V \sim \left| J \right|$, so that capturing a pseudotransition in proximity to the critical point proves to be challenging in practice. However, in extreme cases where $V$ significantly outweighs $J$ or vice versa, the pseudotransition becomes apparent from the side of the CO or (A)FM phases, respectively.

\begin{figure}
	\centering
		\includegraphics[width=\linewidth]{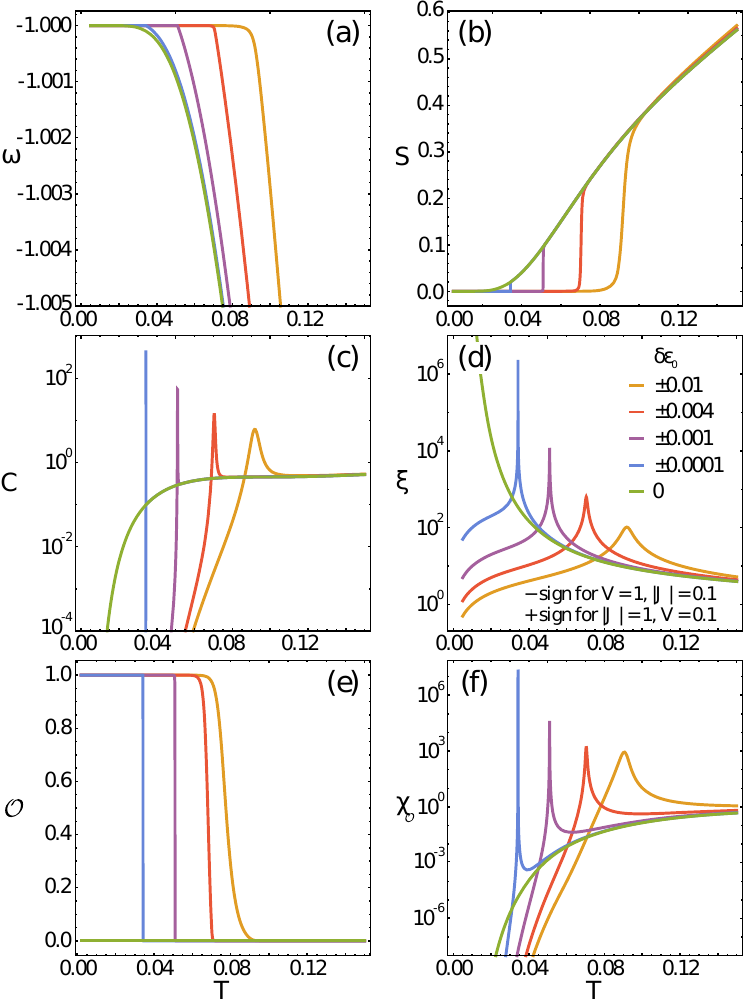}
	\caption{Temperature dependencies of the (a) grand potential, (b) entropy, (c) specific heat, (d) correlation length, (e) order parameter, and (f) susceptibility for $n=0$ and various values of $\Delta$ exhibit typical behavior characteristic of the ``first-order'' pseudotransitions}
	\label{ps1}
\end{figure}

This will be demonstrated by examining the temperature dependencies of thermodynamic quantities. Our findings reveal that the pseudotransitions are identical in the two symmetric regions, $V \gg \left| J \right|$ and $\left| J \right| \gg V$, for $\delta \varepsilon_{0} > 0$ and $\delta \varepsilon_{0} < 0$, respectively. 
For instance, consider two symmetrical situations, with $V = 10 \left| J \right| = 1$ and $\left| J \right| = 10 V = 1$. 
This scenario is illustrated in Fig.~\ref{ps1}, which displays the temperature dependencies of the grand potential, entropy, specific heat, correlation length, order parameter, and corresponding susceptibility for different values of $\delta \varepsilon_{0}$, but one must take $\delta \varepsilon_{0} > 0$ for $V \gg \left| J \right|$, and $\delta \varepsilon_{0} < 0$ for $V \ll \left| J \right|$. 
The curves will be identical for these two ratios. 

Furthermore, for $\delta \varepsilon_{0} > 0$ the order parameter $\mathcal{O}$ will be $M_\text{CO}$. 
In contrast, for $\delta \varepsilon_{0} < 0$, it is necessary to use $m_\text{FM}$ for $J < 0$ and $m_\text{AFM}$ for $J > 0$. 
It is important to note that for the FM case, the system must be considered in a small magnetic field $h \rightarrow 0$ (without loss of generality, we have set $h = 10^{-5}$) to provide the magnetically biased phase. 
Otherwise, the magnetization would be zero when $h=0$.

A high-entropy state at the boundary is characterized by a smooth decrease in entropy to zero with temperature [as illustrated by the green curve in Fig.~\ref{ps1}(b)). 
States in close proximity to the boundary exhibit a sudden transition to a low-entropy state, which becomes increasingly sharp as it approaches the phase boundary. This results in a narrow pseudotransition region, where the grand potential shows a breakpoint, and both entropy and order parameter exhibit sudden jumps, which resemble a typical first-order phase transition. Nevertheless, there are no discontinuities, and all quantities remain as smooth, analytical functions. During this pseudotransition, a transition takes place to the quasiphase that will become the ground state at $T=0$ under these parameters. Moving farther away from the phase boundary causes the pseudotransition to occur at higher temperatures, with the grand potential and entropy becoming showing a more gradual change.

The temperature dependencies of the specific heat, correlation length, and magnetic susceptibility under the same conditions are displayed in Figs.~\ref{ps1}(c,d,f) on a semilogarithmic scale. 
The expression for the correlation length is generally given by~\cite{Rojas2019PRE}:
\begin{equation}
	\xi^{-1} = \ln \left| \frac{\lambda_1}{\lambda_2} \right|.
\end{equation}

Distinct sharp peaks are clearly observed during the pseudotransition, resembling characteristics of a second-order phase transition. However, these peaks do not exhibit divergence, as they remain large yet finite. Moving away from the phase boundary results in broader and reduced peaks.

The expression for pseudocritical temperature can be derived from the square root part of transfer-matrix eigenvalues. Assuming $h=0$ and $\mu = 0$, we can get the expressions for $\lambda_1$ and $\lambda_2$, similarly to those defined in~\cite{Souza2018SSC}:
\begin{equation}
	\lambda_{1,2} = \frac{1}{2} \left(w_1 + w_2 \pm \sqrt{\left(w_1 - w_2 \right)^2 + 4 w_0^2}\right),
\end{equation}
where
\begin{equation}
		w_1 = 2 \cosh \frac{J}{T}, \quad
		w_2 = 2 e^{-\frac{\Delta}{T}} \cosh \frac{V}{T},\quad
		w_0 = e^{-\frac{\Delta}{2 T}}.
\end{equation}
The pseudotransition point is determined by the condition $w_1 = w_2 $. 
In this case, the eigenvalues $\lambda_1$ and $\lambda_2$ will converge as close as possible. 
Consequently, we arrive at a transcendental equation that defines the temperature of the pseudotransition:
\begin{equation}
	\frac{\Delta}{T_p} = \ln \frac{\cosh \left(V/T_{p}\right)}{\cosh \left(J/T_{p}\right)}.
	\label{eq:Tp1}
\end{equation}
It is apparent from this expression that there is no pseudotransition for $V=\left| J \right|$.

\begin{figure}
	\centering
	\includegraphics[width=\linewidth]{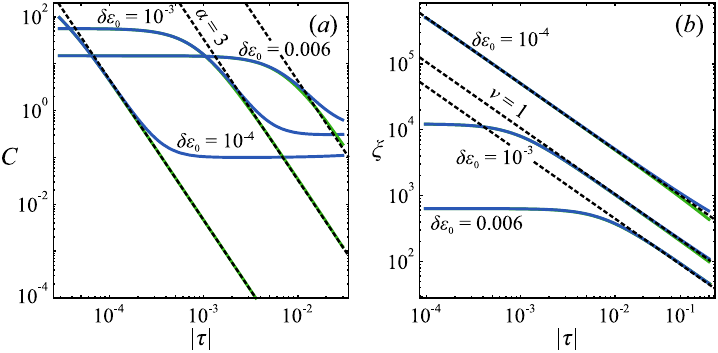}
	\caption{Dependencies of the specific heat (left) and correlation length (right) on $\tau = T - T_p$ shown on a logarithmic scale. Dashed lines represent power-law functions $\ln C(\tau) \sim - 3 \ln \left|\tau\right|$ for specific heat and $\ln \xi(\tau) \sim - \ln \left| \tau \right|$ for correlation length, which match the pseudocritical exponents $\alpha = 3$ and $\nu = 1$. 
		Solid lines correspond to quantities for $\tau > 0$ (blue) and $\tau < 0$ (green)}
	\label{cxi(tau)}
\end{figure}

If the residual entropy experiences a jump discontinuity at the boundary between phases $a$ and $b$, there is the known expression~\cite{Rojas2019PRE} for the pseudotransition temperature $T_p$:
\begin{equation}
	E_a - E_b = T_p (S_a - S_b).
\end{equation}
Thus, near the triple point CO--FM--AFM, on the line $J = 0$, the analytical expression for $T_p$ is provided by the following expression:
\begin{equation}
	T_p = \frac{V-\Delta}{\ln 2}.
	\label{Tp}
\end{equation}
Fig.~\ref{cxi(tau)} displays the temperature dependencies of the specific heat and correlation length on a logarithmic scale in the vicinity of the pseudocritical temperature $\tau = T-T_p$ for the same set of parameters. 
It is apparent that there is an intermediate temperature range that is sufficiently close (but not too close) to a pseudocritical temperature. In this range, both the specific heat and the correlation length exhibit power-law behavior with the pseudocritical exponents $\alpha = \alpha{'} = 3$ and $\nu = \nu{'} = 1$, respectively. 
This is supported by the straight dashed lines observed on the log-log scale in Fig.~\ref{cxi(tau)}. 
Additionally, the magnetic susceptibility also displays a power-law relationship in this temperature range, with pseudocritical exponents $\gamma = \gamma{'} = 3$.

However, it should be mentioned that the correlation length follows the relevant power-law function over wide temperature range symmetrically around $\tau = 0$. 
Specifically, this region is broader for the low-temperature branches ($\tau < 0$) compared to the high-temperature branches ($\tau > 0$) for both specific heat and magnetic susceptibility. Nevertheless, we can conclude that obtained exponents $\alpha = 3$, $\gamma = 3$, $\nu = 1$ satisfy the same universality class found previously for other one-dimensional models; see, for example,~\cite{Rojas2019PRE,Rojas2020BJP}.

Thus, the observed pseudocritical behavior occurs in a ``pure'' case, $n=0$, near the boundary between charge and magnetic orders. 
This finding is unsurprising, as these two phases do not mix with each other. A transition between nonidentical quasiphases with a sharp decrease in entropy is evident. First-order derivatives of the grand potential, including entropy and order parameter, exhibit first-order jumps resembling discontinuities. Second-order derivatives, such as specific heat and susceptibility, manifest sharp peaks akin to second-order phase transitions, yet without any singularities. This pseudotransition with universal pseudocritical exponents has been described in the literature for various one-dimensional and quasi-one-dimensional systems. In addition, the work~\cite{Strecka2024EPJB} revealed this type of pseudotransitions in a similar one-dimensional model for cuprates under weak spin exchange conditions. The electroneutrality constrain imposed in this research is analogous to the ``pure'' state, $n=0$, in the grand canonical formalism employed in our study. 

\subsection{``Second-order'' pseudotransitions}

Now, let us consider the dilute case, $n \neq 0$. 
As mentioned previously, a jump discontinuity in residual entropy is observed at the boundary between the dilute CO and dilute (A)FM phases. The dilute CO phase exhibits high entropy $\mathcal{S}(n)$, while the dilute (A)FM phase with phase separation shows zero residual entropy. Let us identify the pseudotransition near this boundary.

Here, in contrast to the case $n=0$,  the inherent symmetry of the spin and pseudospin systems is broken. Now a pseudotransition can exclusively occur only from the (A)FM side, as $\left|J\right| > V$. 
Conversely, the region in the GSPD [see Figs.~\ref{GSPD}(b,c)] corresponding to the CO phase, with $\left|J\right| < V$, is substituted by the PM-CO phase for $ 0 < \left|n\right| < 1/2$, or FR-PM phase for $\left|n\right| \geq 1/2$, whereas the region of the (A)FM phase is occupied by the FR(A)FM and FR-PM phases, respectively.
Consequently, the boundary $\Delta = V - \left|J\right|$ does not satisfy the Rojas criterion for $\left|J\right| < V$ in a dilute case.

\begin{figure}
	\centering
		\includegraphics[width=\linewidth]{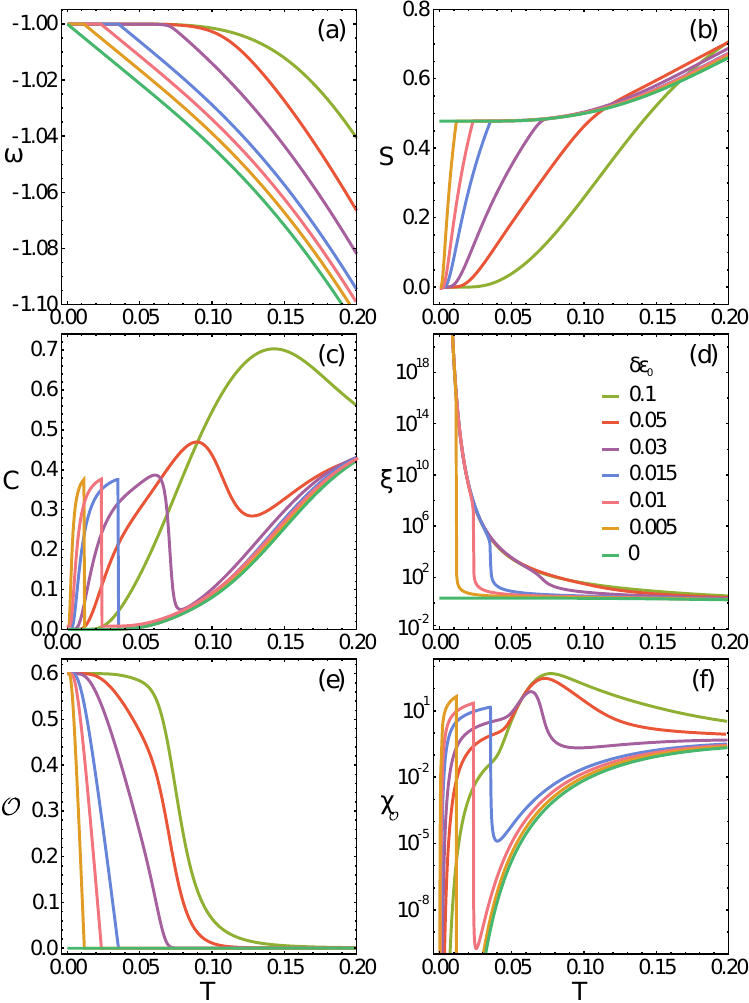}
	\caption{Temperature dependencies of the (a) grand potential, (b) entropy, (c) specific heat, (d) correlation length, (e) order parameter, and (f) susceptibility for $n=0.4$, $V=1/10$, $\left|J\right| = 1$, and various values of $\delta \varepsilon_0$. 
	Significantly different behavior compared to previous pseudotransitions resembling second-order phase transitions is observed.}
	\label{ps2}
\end{figure}

\begin{figure}
	\centering
	\includegraphics[width=\linewidth]{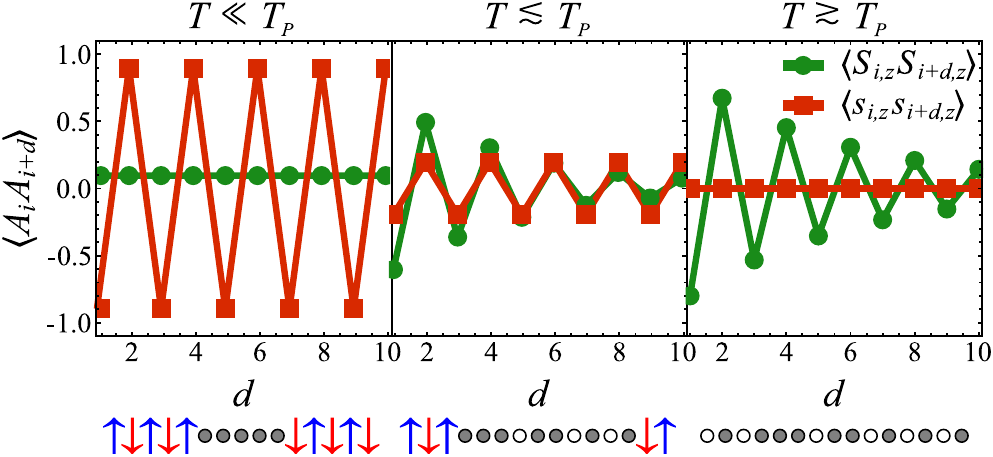}
	\caption{Schematic illustrations depicting the changes in the chain state during the ``second-order'' pseudotransition, $n = 0.1$, $\delta \varepsilon_0 = 0.002$, $J=1$. 
	Green circles denote the density-density correlator $\langle S_{i,z} S_{i+d,z} \rangle$; red squares correspond to the spin correlator $\langle s_{i,z} s_{i+d,z} \rangle$. 
	The distance between nodes is labeled as $d = j-i$. Gray dots (\psup) and white dots (\psdown) denote the two charge states of pseudospin, with $S_z = \pm 1$. 
	Blue (\spinup) and red (\spindown) arrows denote the two magnetic states of spin, with $s_z = \pm 1/2$, which correspond to $S_z = 0$. 
	The system transitions from a dilute CO quasiphase to an AFM quasiphase with short-range charge-ordered droplet. As the temperature decreases, the droplet becomes homogeneous.}
	\label{ps2-chain}
\end{figure}

In Fig.~\ref{ps2}the temperature dependencies of the grand potential, entropy, specific heat, correlation length, order parameter, and corresponding susceptibility are presented for $n=0.4$ and various values of $\delta \varepsilon_0 > 0$, as we approach the CO-(A)FM boundary from the dilute (A)FM side, and $\left|J\right| > V$.
The grand potential displays nearly linear behavior up to the pseudotransition temperature, characterized by a sudden breakpoint. Subsequently, at low temperatures, the grand potential stabilizes at a near-constant value. The entropy decreases gradually at higher temperatures, eventually reaching a plateau where it remains nearly constant, corresponding to the residual entropy of the CO phase. It can be inferred that the system reaches the frustrated CO quasiphase. A clear pseudocritical breakpoint in the entropy is then observed, indicating the beginning of a gradual decrease. Nevertheless, the entropy still remains continuous everywhere. A pseudotransition from the frustrated CO quasiphase to a dilute (A)FM quasiphase takes place. The changes in the chain states during this process can be observed in Fig.~\ref{ps2-chain} for the AFM scenario. Interestingly, the droplet in a dilute magnetic quasiphase contains a short-range charge order, resulting in a nonzero entropy. As the temperature decreases, the droplet becomes more homogeneous until its volume matches $\left| n \right|$, and the entropy reaches zero. 
Additionally, following the pseudotransition, the order parameter smoothly attains its maximum value of $1-\left|n\right|$, as illustrated in Fig.~\ref{ps2}(e).


Examine the temperature dependencies of the specific heat, correlation length, and magnetic susceptibility depicted in Fig.~\ref{ps2}(c,d,f). 
Near the border $\Delta = V - \left| J \right|$ ($\delta \varepsilon_0 = 0$), the specific heat and susceptibility have two minor peaks. The first peak, at high temperatures, indicates the formation of a short-range charge-ordered quasiphase. As the entropy plateaus, the specific heat and susceptibility reach minimum values, which indicates that the dilute CO quasiphase is almost frozen. At the pseudotransition point, a finite sudden jump is observed, although of small magnitude, without any discontinuity. This behavior closely resembles that of a second-order phase transition. Subsequently, the specific heat and magnetic susceptibility gradually decrease to zero. As one moves farther away from the boundary, the region with the jump widens, and the pseudotransition occurs more smoothly and at higher temperatures. Consequently, at a significant distance from the boundary, the pseudotransition vanishes, and the quasiphase exhibits magnetic order with a charge droplet, which is already homogeneous. The behavior of the correlation length is notable, showing a significant abrupt increase at $T=T_p$, as shown in Fig.~\ref{ps2}(d), followed by divergence as $T$ approaches zero.

Given that this anomalous temperature behavior bears resemblance to a referential second-order phase transition, characterized by discontinuities in the second derivatives of the grand potential (the specific heat, correlation length, and magnetic susceptibility), we designate this discovered pseudotransition as a ``second-order'' pseudotransition. Following the analogy with Ehrenfest classification, the transitions discussed in the previous section were termed ``first-order'' pseudotransitions, as indicated in the section heading. It is important to note, however, that this classification may not be entirely suitable. These ``first-order'' pseudotransitions exhibit almost discontinuities in both the first and second derivatives of the grand potential, along with universal critical exponents $\alpha$, $\gamma$ and $\nu$.
In essence, ``first-order'' pseudotransitions display characteristics of both first-order and second-order phase transitions. On the other hand, ``second-order'' pseudotransitions lack the critical exponents typically associated with second-order phase transitions. Additionally, the phenomenological explanation of this pseudotransition is centered around nucleation, a characteristic more commonly observed in first-order phase transitions.

\subsection{Phenomenology of ``second-order'' pseudotransitions}

To determine the temperatures of ``second-order'' pseudotransitions, we will employ the Maxwell construction~\cite{Plischke2006}. 
Assuming the presence of two coexisting macroscopic homogeneous phases 1 and 2, the free energy of the phase separation state per node is expressed as
\begin{equation}
	f_{\text{PS} } = p f_1 (n_1) + (1 - p) f_2 (n_2), 
	\label{eq:fPS}
\end{equation}
where $p$ is the system fraction with density $n_1$, $1 - p$ is the system fraction with density $n_2$, so that $p n_1 + (1 - p) n_2 = n$. 
In our case, the first phase is an (A)FM nucleus, which arises in the pseudotransition. 
The second phase is a dilute CO quasi-phase, which is formed at $T \gtrsim T_p$, when the entropy reaches plateau. 
Therefore, $n_1 = 0$, and $n_2 = n +\delta n$, since the CO state must has an increasing charge density to compensate for the decrease in volume due to the nucleation of the (A)FM phase. If $p=0$, the equality $f_\text{PS} = f_\text{CO} (n)$ holds. Assuming that in a small vicinity below the pseudotransition point $p \rightarrow 0$, $\delta n = p \cdot n \rightarrow 0$, we use the first variation of Eq.~\eqref{eq:fPS} to obtain 
\begin{equation}
	f_\text{(A)FM} (0) -  f_\text{CO} (n) + n \, f'_\text{CO} (n) = 0 . 
\end{equation}
The free energies of (A)FM and CO phases we can write as
\begin{eqnarray}
	f_\text{(A)FM} (0) & \simeq & \varepsilon_\text{(A)FM}^{0} (0) , 
	\label{eq:free0AFM}
	\\
	f_\text{CO} (n) & \simeq & \varepsilon_\text{CO}^{0} (n) - T \mathcal{S}(n) . 
	\label{eq:free0CO}
\end{eqnarray}
The index 0 indicates $T=0$. 
Here we posit that the CO quasi-phase is practically stationary for $T \gtrsim T_p$, with energy and entropy nearly identical to those in the ground state, so $\mathcal{S}(n)$ is defined by Eq.~\eqref{eq:S(n)}. 
We also assume that during a pseudotransition, an almost ordered (A)FM phase is formed, characterized by a sufficiently large correlation length, as actually shown in Fig.~\ref{ps2}(d). 
Thus, we obtain the following expression for the pseudocritical temperature:
\begin{equation}
	T_{p,2} 
	= \frac{ 2\, \delta \varepsilon_{0} }{\ln\left(1+\left|n\right|\right) - \ln\left(1-\left|n\right|\right)} .
	\label{eq:Tp(n)}
\end{equation}

\begin{figure}
	\centering
	\includegraphics[width=0.75\linewidth]{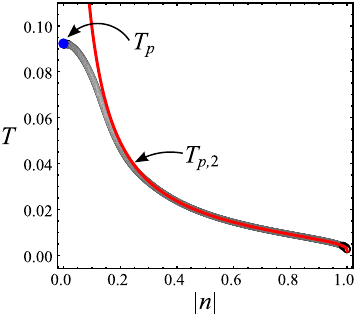}
	\caption{The dependency of pseudocritical temperature on charge density $n$ for $\delta \varepsilon_0 =  0.01$. 
	The black circles represent calculations based on the maximum of the specific heat, while the red solid line corresponds to the phenomenological estimation for pseudocritical temperature~\eqref{eq:Tp(n)}. 
	The blue dot is in accordance with Eq.~\eqref{eq:Tp1} for the ``first-order'' pseudotransition}
	\label{fig:Tp(n)}
\end{figure}

\begin{figure*}
	\centering
	\includegraphics[width=\linewidth]{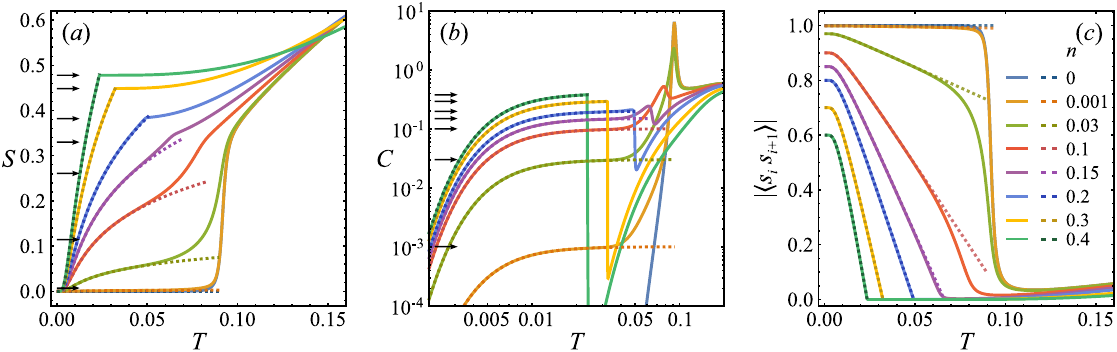}
	\caption{Temperature dependencies of the (a) entropy, (b) specific heat, and (c) absolute value of the spin correlator between two neighboring sites for various values of charge density $n$. The solid colored lines represent results from exact calculations, while the dashed colored lines depict the estimations for the fraction of the (A)FM phase given by Eq.~\eqref{eq:p(A)FM}. The black arrows correspond to (a) the residual entropy of CO phase given by Eq.~\eqref{s(n)} and (b) the estimation of the specific heat jump given by Eq.~\eqref{eq:dC} }
	\label{fig:scp(T)-n}
\end{figure*}

Figure~\ref{fig:Tp(n)} displays the concentration dependency of the pseudocritical temperature for $\delta \varepsilon_0 = 0.01$. 
It is evident that in the high-concentration range, the dependency closely aligns with equation~\eqref{eq:Tp(n)}, 
since our assumption~\eqref{eq:free0CO} of a weak dependency of entropy on $T$ is well confirmed for the high-temperature CO quasi-phase.
However, for small values of $n$, this approximation ceases to be valid. 
Calculated entropy dependencies in Fig.~\ref{fig:scp(T)-n}(a) reveal the reason of this deviation: 
the entropy of the high-temperature CO quasi-phase depends significantly on $T$ for $|n|<0.2$, and this is not taken into account when obtaining expression~\eqref{eq:Tp(n)}. 
Horizontal dashed lines in Fig.~\ref{fig:scp(T)-n}(a) indicate the residual entropy $\mathcal{S}(n)$ of the dilute CO phase~\eqref{eq:S(n)}, and if $|n|\ge0.2$, the entropy dependencies have plateau at the level of $\mathcal{S}(n)$. 
This is what is assumed in expression~\eqref{eq:free0CO} and makes estimation~\eqref{eq:Tp(n)} for $T_{p,2}$ appropriate.

In the limit of extremely small $n$, the dependency of pseudocritical temperature on $n$ in Fig.~\ref{fig:Tp(n)} reaches a value equal to the pseudocritical temperature for the ``first-order'' pseudotransition, determined by equation~\eqref{eq:Tp1}.
Moreover, the thermodynamic functions also retain features of the ``first-order'' pseudotransition at low charge densities. 
Fig.~\ref{fig:scp(T)-n} illustrates the temperature dependencies of the entropy, specific heat, and absolute value of the spin correlator between two neighboring sites for $\delta \varepsilon_0 = 0.01$ and various values of $n$. 
The dependencies of the specific heat for $n=0, \; 0.001, \; 0.03$ show similar peaks at $T=T_{p}$, with the peak height decreasing with the growth of $n$. 
At the same time, the drop in entropy at $T>T_{p}$ for these $n$ still remains significant, similar to the observations made for the ``first-order'' pseudotransitions in Fig.~\ref{ps1}(b). 
There is also an absence of a plateau at $T \gtrsim T_p$ that would correspond to the residual entropy of the CO phase. 
The corresponding values of $\mathcal{S}(n)$ are depicted with black arrows in Fig.~\ref{fig:scp(T)-n}(a).
The dependencies of the specific heat for $n =0.1$ and $0.15$ show a smooth transition to the dependencies typical of the ``second-order'' pseudotransitions.
It can be seen that the formation of a plateau in the temperature dependency of entropy at $T\gtrsim T_{p}$ leads to a drop in the specific heat, and its peak smoothly evolves into the edge of a sharp rise at $T=T_{p}$.

The continuous shift from a ``first-order'' to a ``second-order'' pseudotransition indicates a shared underlying mechanism for these phenomena.
Further insights are revealed from the concentration dependencies of the spin pair distribution function for nearest neighbors shown in Fig.~\ref{fig:scp(T)-n}(c).
This function represents the combination of probabilities associated with four possible pairs of spin states~\cite{Panov2020JMMM}. 
But at the chosen parameters and low temperatures, it is equal to the total probability of pairs with (anti)ferromagnetic spin ordering. Due to the unique feature of the (A)FM phase, which remains undiluted by impurities and does not mix with the CO phase, the $\left|\left\langle s_{i}s_{i+1}\right\rangle\right|$ value provides a reliable estimate of the fraction $p$ of this phase in the chain. As depicted in  Fig.~\ref{fig:scp(T)-n}(c), the fraction of the (A)FM phase at $n=0$ experiences a sudden change from 0 to 1 during the pseudotransition.
Conversely, for $n>0.15$, this fraction increases linearly from 0 to a maximum possible value of $1-\left|n\right|$ as temperature decreases.
The dependencies for $0<n<0.15$ exhibit an intermediate behavior.

It is natural to use the same simple phenomenological concept of phase separation to explain the temperature dependency of the fraction of (A)FM phase, similar to the derivation of Eq.~\eqref{eq:Tp(n)}, and compare this with the exact result. 
By taking the derivative of $f_\text{PS}$ in Eq.\eqref{eq:fPS} with respect to temperature, we can determine the entropy of the system, which aligns with the entropy of the dilute CO phase at  $T<T_{p,2}$:
\begin{equation}
	\frac{\partial f_\text{PS}}{\partial T} = - \mathcal{S}_\text{PS} = - \left(1-p\right) \mathcal{S}(n_\text{CO}) , 
\end{equation}
where $ n_\text{CO} = n/\left(1-p\right)$. 
The derivative of the right-hand side of Eq.~\eqref{eq:fPS}, with taking into account~\eqref{eq:free0AFM} and~\eqref{eq:free0CO}, can be written in the following form:
\begin{multline}
	\frac{\partial}{\partial T} \left[ p \, f_\text{(A)FM}(0) 
	+ \left(1-p\right) f_\text{CO}(n_\text{CO})\right] 
	= - \delta \varepsilon_{0} \cdot \frac{\partial p}{\partial T} \\
	- \left(1-p\right) \mathcal{S}(n_\text{CO}) 
	- T \frac{\partial}{\partial T} \left[\left(1-p\right) \mathcal{S}(n_\text{CO})\right] , 
\end{multline}
where
\begin{equation}
	\frac{\partial}{\partial T} \left[\left(1-p\right) \mathcal{S}(n_\text{CO})\right] 
	= - \frac{1}{2} \ln \left(\frac{1-p+|n|}{1-p-|n|}\right) \, \frac{\partial p}{\partial T} . 
\end{equation}
Equating the derivatives of the right and left parts of Eq.~\eqref{eq:fPS}, we obtain an expression for the fraction of (A)FM phase:
\begin{equation}
	p = 1 - |n| \, \coth \left( \beta \, \delta \varepsilon_{0} \right) . 
	\label{eq:p(A)FM} 
\end{equation}
This equation gives $p=0$ at $T=T_{p,2}$, and $p=1-|n|$ at $T=0$.
A comparison of the dependency~\eqref{eq:p(A)FM} with the exact result is shown in Fig.~\ref{fig:scp(T)-n}(c) with dashed lines. 
For $0<n<0.15$ the fitting is quite close, and even for $n=0.1$ and $n=0.03$ the low-temperature part of the curves is described by Eq.~\eqref{eq:p(A)FM}.

It is the temperature dependency of $p$ that causes the entropy of the system to change in the phase separation state, so Eq.~\eqref{eq:p(A)FM} defines the temperature dependency at $T<T_{p,2}$ in phenomenological expressions for entropy:
\begin{multline}
	\mathcal{S}_{PS} = \left(1-p\right) \mathcal{S}\left(\frac{n}{1-p}\right) \\
	= |n| \Big\{ \beta\,\delta \varepsilon_{0} \coth \left(\beta\,\delta\varepsilon_{0}\right) 
	- \ln \left[2 \sinh \left(\beta \, \delta \varepsilon_{0}\right)\right] \Big\} , 
	\label{eq:S_PS} 
\end{multline}
and for the specific heat:
\begin{equation}
	C_\text{PS} = - \delta \varepsilon_{0} \cdot \frac{\partial p}{\partial T} 
	= |n| \left[\frac{\beta\,\delta\varepsilon_{0}}{\sinh\left(\beta\,\delta\varepsilon_{0}\right)}\right]^{2} . 
	\label{eq:C_PS}
\end{equation}
These expressions are in excellent agreement with exact results for the entropy and specific heat in the low-temperature region, as illustrated by the dashed colored lines in Figs.~\ref{fig:scp(T)-n}(a,b).

\begin{figure}
	\centering
	\includegraphics[width=0.75\linewidth]{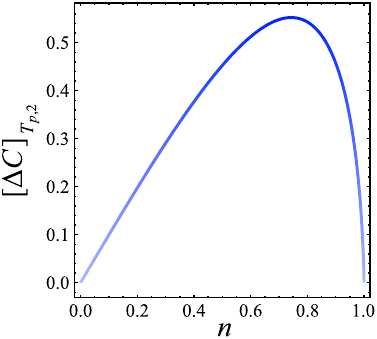}
	\caption{The dependency of the specific heat jump at $T=T_{p,2}$ on charge density $n$ given by Eq.~\eqref{eq:dC}  }
	\label{fig:dC}
\end{figure}

Finally, the combined use of Eqs.~\eqref{eq:Tp(n)} and~\eqref{eq:C_PS} enables us to obtain the jump in the specific heat at $T=T_{p,2}$, which has a universal form and depends only on the charge density $n$:
\begin{equation}
	\left[\Delta C \right]_{T_{p,2}} = \frac{1-n^{2}}{4|n|} \left(\ln\frac{1+|n|}{1-|n|}\right)^{2} . 
	\label{eq:dC}
\end{equation}
This function is shown in Fig.~\ref{fig:dC}.
The maximum of the specific heat jump $\Delta C_{max}\simeq0.55$ corresponds to $n\simeq0.74$.
This specific value of $n$ results from a compromise between the tendency for the pseudo-transition temperature to decrease and the sharper slope of the $\mathcal{S}_{\text{PS}}$ curve as $\left| n \right|$ increases.

\section{Conclusions}
In the present article, we have addressed the study of a one-dimensional spin chain diluted by two types of charged nonmagnetic impurities. The peculiarity of the model is the condition of constant charge density of the system $n$ with a possible variable number of impurities of each type and spin centers. To accurately calculate thermodynamic averages, we introduce a high-performance method that utilizes the properties of the transfer-matrix of the system.

Assuming a fixed charge density, the ground state phase diagram includes phases that do not form a mixed state at the phase boundaries. According to the Rojas criterion~\cite{Rojas2020BJP,*Rojas2020APPA}, pseudotransitions can be observed near these boundaries. These pseudotransitions are characterized by striking features of the specific heat, correlation length, order parameter, and susceptibility at some finite temperature. Thus, at $n=0$ near the boundary between CO and (A)FM phases, pseudotransitions are observed both in the CO phase region for $V>\left|J\right|$, as was previously discussed in~\cite{Strecka2024EPJB}, and in the AFM phase region for $V<\left|J\right|$. 
This highlights the inherent symmetry of the spin and pseudospin systems. These pseudotransitions exhibit known features, such as characteristic values of pseudocritical exponents, and are referred to as ``first-order'' pseudotransitions.

At $n\neq0$, the ``second-order'' pseudotransitions are observed in the system. These are characterized by a distinct breakpoint in the temperature dependency of the entropy and order parameter, which is different from a stepwise behavior observed in the ``first-order'' pseudotransitions. Furthermore, they exhibit sharp, finite jumps in the specific heat and magnetic susceptibility, akin to the behavior observed in conventional second-order phase transitions.

We show that there is a continuous transformation of the type of pseudotransitions, transitioning from the ``first-order'' at 
$n
=
0$
to the ``second-order'' as 
$|
n
|$
grows. The specificity of CO and (A)FM states to alternate during the pseudotransition allows us to plot the temperature dependency of the fraction of the (A)FM phase. The pseudotransition at 
$n
=
0$
is characterized by an abrupt change in the state of the entire chain. In the ``second-order'' pseudotransition, the (A)FM phase appears in the form of nuclei, and beyond 
$|
n
|
>
0.15$, the fraction of the (A)FM phase increases linearly with decreasing temperature. In this case, it is possible to describe with high accuracy the thermodynamic properties of the system in the framework of a simple phenomenological Maxwell construction for two quasiphases, dilute CO and (A)FM. This enables us to find phenomenological dependencies of the ``second-order'' pseudotransition temperature and the fraction of the (A)FM phase, which, for 
$|
n
|
>
0.15$, completely determines the behavior of the system below the pseudotransition point. (A)FM phase replaces the dilute CO phase with decreasing temperature, and the resulting increase in charge density in the dilute CO phase explains the temperature dependency of the entropy and specific heat of the whole system. The inapplicability of the phenomenological description for 
$|
n
|
<
0.15$
is attributed to the absence of the dilute CO quasiphase formation up to the pseudotransition point, as evident from the temperature dependency of the system's entropy.

A comprehensive examination of the one-dimensional model facilitates a comparison between accurate calculations and reasonable phenomenological constructions, aiding in the comprehension of the peculiarities of physical properties and the limits of phenomenological applicability. This allows us to use this experience in more complex systems exhibiting phase separation.

\section{Acknowledgments}

The research funding from the Ministry of Science and Higher Education of the Russian Federation (Ural Federal University Program of Development within the Priority-2030 Program) is gratefully acknowledged. D.Y. acknowledges funding by the Theoretical Physics and Mathematics Advancement Foundation ``BASIS'', Grant No. 22-1-5-123-1.

\appendix
\section{The expression for eigenvalues}

Eigenvalues $\{\lambda_i, \, i=1, \ldots 4\}$ of the transfer-matrix~(\ref{TM}) are determined by solving the equation
\begin{equation}
	\vert \hat{T} - \lambda \hat{I} \vert = \lambda^4 + a \lambda^3 + b \lambda^2 + c \lambda + d = 0,
	\label{lambda_eq}
\end{equation}
where $\hat{I}$ is an identity matrix,
\begin{eqnarray}
	\!\!\! a &=& -2 e^{-\frac{J}{T}}-2 e^{-\frac{\Delta +V}{T}} \cosh \frac{\mu }{T} , \\ 
	\!\!\! b &=& -4 e^{-\frac{\Delta+V+J}{T}} \left( e^{\frac{V+J}{T}} - 1 \right)  \cosh \frac{\mu}{T} \\ 
	\!\!\! &&- 2 \sinh \frac{2J}{T} -2 e^{-\frac{2\Delta}{T}} \sinh \frac{2V}{T} , \\ 
	\!\!\! c &=& 8 e^{-\frac{\Delta}{T}} \sinh \frac{J}{T} \cosh \frac{\mu}{T} \left( e^{-\frac{V}{T}} \cosh \frac{J}{T} -1 \right) \\ 
	\!\!\! &&{}+8 e^{-\frac{2\Delta}{T}} \sinh \frac{V}{T} \left( e^{-\frac{J}{T}} \cosh \frac{V}{T} -1 \right) \! , \\ 
	\!\!\! d &=& 16 e^{-\frac{2\Delta}{T}} \!\! \left( \! \cosh \frac{V}{T} \cosh \frac{J}{T}  -1 \! \right) \sinh \frac{V}{T} \sinh \frac{J}{T} . 
	\label{coef}
\end{eqnarray}

Solutions of equation~\eqref{lambda_eq} for eigenvalues of transfer-matrix can be derived using Ferrari's method~\cite{Korn2000}:
\begin{equation}
	\lambda_{1,2,3,4} = -\frac{a}{4} + \frac{\pm_s \sqrt{\alpha+2y} \pm_t \sqrt{-(3 \alpha + 2y \pm_s \frac{2 \beta}{\sqrt{\alpha+2y}})}}{2},
\end{equation}
where 
\begin{eqnarray}
	\alpha &=& - \frac{3a^2}{8} + b ,\\
	\beta &=& \frac{a^3}{8} - \frac{a b}{2} + c ,\\
	\gamma &=& - \frac{3 a^4}{256} + \frac{a^2 b}{16} - \frac{a c}{4} + d,\\
	p &=& -\frac{\alpha^2}{12} - \gamma,\\
	q &=& - \frac{\alpha^3}{108} + \frac{\alpha \gamma}{3} - \frac{\beta^2}{8} ,\\
	r &=& - \frac{q}{2} + \sqrt{\frac{q^2}{4}+\frac{p^3}{27}} ,\\
	u &=& \sqrt[3]{r} ,\\
	y &=& -\frac{5}{6} \alpha + u +
	\begin{cases}
		-\sqrt[3]{q}, & u = 0 ; \\
		- \frac{p}{3u}, & u \neq 0 . \\
	\end{cases}
\end{eqnarray}

Here $\pm_s$ and $\pm_t$ are two independent parameters, each of which is equal to either $+$ or $-$. The number of possible pairs of their values is four, and each pair produces one of the four eigenvalues. The maximum eigenvalue $\lambda_1$ corresponds to $\pm_s = \pm_t = +$.

\bibliographystyle{apsrev4-2}
\bibliography{Manuscript_PRE2024}

\begin{thebibliography}{57}%
\makeatletter
\providecommand \@ifxundefined [1]{%
 \@ifx{#1\undefined}
}%
\providecommand \@ifnum [1]{%
 \ifnum #1\expandafter \@firstoftwo
 \else \expandafter \@secondoftwo
 \fi
}%
\providecommand \@ifx [1]{%
 \ifx #1\expandafter \@firstoftwo
 \else \expandafter \@secondoftwo
 \fi
}%
\providecommand \natexlab [1]{#1}%
\providecommand \enquote  [1]{``#1''}%
\providecommand \bibnamefont  [1]{#1}%
\providecommand \bibfnamefont [1]{#1}%
\providecommand \citenamefont [1]{#1}%
\providecommand \href@noop [0]{\@secondoftwo}%
\providecommand \href [0]{\begingroup \@sanitize@url \@href}%
\providecommand \@href[1]{\@@startlink{#1}\@@href}%
\providecommand \@@href[1]{\endgroup#1\@@endlink}%
\providecommand \@sanitize@url [0]{\catcode `\\12\catcode `\$12\catcode
  `\&12\catcode `\#12\catcode `\^12\catcode `\_12\catcode `\%12\relax}%
\providecommand \@@startlink[1]{}%
\providecommand \@@endlink[0]{}%
\providecommand \url  [0]{\begingroup\@sanitize@url \@url }%
\providecommand \@url [1]{\endgroup\@href {#1}{\urlprefix }}%
\providecommand \urlprefix  [0]{URL }%
\providecommand \Eprint [0]{\href }%
\providecommand \doibase [0]{https://doi.org/}%
\providecommand \selectlanguage [0]{\@gobble}%
\providecommand \bibinfo  [0]{\@secondoftwo}%
\providecommand \bibfield  [0]{\@secondoftwo}%
\providecommand \translation [1]{[#1]}%
\providecommand \BibitemOpen [0]{}%
\providecommand \bibitemStop [0]{}%
\providecommand \bibitemNoStop [0]{.\EOS\space}%
\providecommand \EOS [0]{\spacefactor3000\relax}%
\providecommand \BibitemShut  [1]{\csname bibitem#1\endcsname}%
\let\auto@bib@innerbib\@empty
\bibitem [{\citenamefont {Van~Hove}(1950)}]{vanHove1950Physica}%
  \BibitemOpen
  \bibfield  {author} {\bibinfo {author} {\bibfnamefont {L.}~\bibnamefont
  {Van~Hove}},\ }\href {https://doi.org/10.1016/0031-8914(50)90072-3}
  {\bibfield  {journal} {\bibinfo  {journal} {Physica}\ }\textbf {\bibinfo
  {volume} {16}},\ \bibinfo {pages} {137} (\bibinfo {year} {1950})}\BibitemShut
  {NoStop}%
\bibitem [{\citenamefont {Cuesta}\ and\ \citenamefont
  {S{\'a}nchez}(2004)}]{Cuesta2004JSP}%
  \BibitemOpen
  \bibfield  {author} {\bibinfo {author} {\bibfnamefont {J.~A.}\ \bibnamefont
  {Cuesta}}\ and\ \bibinfo {author} {\bibfnamefont {A.}~\bibnamefont
  {S{\'a}nchez}},\ }\href {https://doi.org/10.1023/B:JOSS.0000022373.63640.4e}
  {\bibfield  {journal} {\bibinfo  {journal} {Journal of Statistical Physics}\
  }\textbf {\bibinfo {volume} {115}},\ \bibinfo {pages} {869} (\bibinfo {year}
  {2004})}\BibitemShut {NoStop}%
\bibitem [{\citenamefont {O{'}Hare}\ \emph {et~al.}(2007)\citenamefont
  {O{'}Hare}, \citenamefont {Kusmartsev}, \citenamefont {Kugel},\ and\
  \citenamefont {Laad}}]{OHare2007Aug}%
  \BibitemOpen
  \bibfield  {author} {\bibinfo {author} {\bibfnamefont {A.}~\bibnamefont
  {O{'}Hare}}, \bibinfo {author} {\bibfnamefont {F.~V.}\ \bibnamefont
  {Kusmartsev}}, \bibinfo {author} {\bibfnamefont {K.~I.}\ \bibnamefont
  {Kugel}},\ and\ \bibinfo {author} {\bibfnamefont {M.~S.}\ \bibnamefont
  {Laad}},\ }\href {https://doi.org/10.1103/PhysRevB.76.064528} {\bibfield
  {journal} {\bibinfo  {journal} {Phys. Rev. B}\ }\textbf {\bibinfo {volume}
  {76}},\ \bibinfo {pages} {064528} (\bibinfo {year} {2007})}\BibitemShut
  {NoStop}%
\bibitem [{\citenamefont {Yin}(2024{\natexlab{a}})}]{Yin2024PRR}%
  \BibitemOpen
  \bibfield  {author} {\bibinfo {author} {\bibfnamefont {W.}~\bibnamefont
  {Yin}},\ }\href {https://doi.org/10.1103/PhysRevResearch.6.013331} {\bibfield
   {journal} {\bibinfo  {journal} {Physical Review Research}\ }\textbf
  {\bibinfo {volume} {6}},\ \bibinfo {pages} {013331} (\bibinfo {year}
  {2024}{\natexlab{a}})}\BibitemShut {NoStop}%
\bibitem [{\citenamefont {Yin}(2024{\natexlab{b}})}]{Yin2024PRB}%
  \BibitemOpen
  \bibfield  {author} {\bibinfo {author} {\bibfnamefont {W.}~\bibnamefont
  {Yin}},\ }\href {https://doi.org/10.1103/PhysRevB.109.214413} {\bibfield
  {journal} {\bibinfo  {journal} {Phys. Rev. B}\ }\textbf {\bibinfo {volume}
  {109}},\ \bibinfo {pages} {214413} (\bibinfo {year}
  {2024}{\natexlab{b}})}\BibitemShut {NoStop}%
\bibitem [{\citenamefont {Rojas}(2020{\natexlab{a}})}]{Rojas2020BJP}%
  \BibitemOpen
  \bibfield  {author} {\bibinfo {author} {\bibfnamefont {O.}~\bibnamefont
  {Rojas}},\ }\href {https://doi.org/10.1007/s13538-020-00773-8} {\bibfield
  {journal} {\bibinfo  {journal} {Brazilian Journal of Physics}\ }\textbf
  {\bibinfo {volume} {50}},\ \bibinfo {pages} {675} (\bibinfo {year}
  {2020}{\natexlab{a}})}\BibitemShut {NoStop}%
\bibitem [{\citenamefont {Rojas}(2020{\natexlab{b}})}]{Rojas2020APPA}%
  \BibitemOpen
  \bibfield  {author} {\bibinfo {author} {\bibfnamefont {O.}~\bibnamefont
  {Rojas}},\ }\href {https://doi.org/10.12693/APhysPolA.137.933} {\bibfield
  {journal} {\bibinfo  {journal} {Acta Physica Polonica A}\ }\textbf {\bibinfo
  {volume} {137}},\ \bibinfo {pages} {933} (\bibinfo {year}
  {2020}{\natexlab{b}})}\BibitemShut {NoStop}%
\bibitem [{\citenamefont {Panov}\ and\ \citenamefont
  {Rojas}(2021)}]{Panov2021PRE}%
  \BibitemOpen
  \bibfield  {author} {\bibinfo {author} {\bibfnamefont {Y.}~\bibnamefont
  {Panov}}\ and\ \bibinfo {author} {\bibfnamefont {O.}~\bibnamefont {Rojas}},\
  }\href {https://doi.org/10.1103/PhysRevE.103.062107} {\bibfield  {journal}
  {\bibinfo  {journal} {Physical Review E}\ }\textbf {\bibinfo {volume}
  {103}},\ \bibinfo {pages} {062107} (\bibinfo {year} {2021})}\BibitemShut
  {NoStop}%
\bibitem [{\citenamefont {Rojas}\ \emph {et~al.}(2019)\citenamefont {Rojas},
  \citenamefont {Stre{\v c}ka}, \citenamefont {Lyra},\ and\ \citenamefont
  {de~Souza}}]{Rojas2019PRE}%
  \BibitemOpen
  \bibfield  {author} {\bibinfo {author} {\bibfnamefont {O.}~\bibnamefont
  {Rojas}}, \bibinfo {author} {\bibfnamefont {J.}~\bibnamefont {Stre{\v c}ka}},
  \bibinfo {author} {\bibfnamefont {M.~L.}\ \bibnamefont {Lyra}},\ and\
  \bibinfo {author} {\bibfnamefont {S.~M.}\ \bibnamefont {de~Souza}},\ }\href
  {https://doi.org/10.1103/PhysRevE.99.042117} {\bibfield  {journal} {\bibinfo
  {journal} {Physical Review E}\ }\textbf {\bibinfo {volume} {99}},\ \bibinfo
  {pages} {042117} (\bibinfo {year} {2019})}\BibitemShut {NoStop}%
\bibitem [{\citenamefont {Krokhmalskii}\ \emph {et~al.}(2021)\citenamefont
  {Krokhmalskii}, \citenamefont {Hutak}, \citenamefont {Rojas}, \citenamefont
  {de~Souza},\ and\ \citenamefont {Derzhko}}]{Krokhmalskii2021PhysicaA}%
  \BibitemOpen
  \bibfield  {author} {\bibinfo {author} {\bibfnamefont {T.}~\bibnamefont
  {Krokhmalskii}}, \bibinfo {author} {\bibfnamefont {T.}~\bibnamefont {Hutak}},
  \bibinfo {author} {\bibfnamefont {O.}~\bibnamefont {Rojas}}, \bibinfo
  {author} {\bibfnamefont {S.~M.}\ \bibnamefont {de~Souza}},\ and\ \bibinfo
  {author} {\bibfnamefont {O.}~\bibnamefont {Derzhko}},\ }\href
  {https://doi.org/10.1016/j.physa.2021.125986} {\bibfield  {journal} {\bibinfo
   {journal} {Physica A}\ }\textbf {\bibinfo {volume} {573}},\ \bibinfo {pages}
  {125986} (\bibinfo {year} {2021})}\BibitemShut {NoStop}%
\bibitem [{\citenamefont {Hovhannisyan}\ \emph {et~al.}(2016)\citenamefont
  {Hovhannisyan}, \citenamefont {Stre{\v c}ka},\ and\ \citenamefont
  {Ananikian}}]{Hovhannisyan2016JPCM}%
  \BibitemOpen
  \bibfield  {author} {\bibinfo {author} {\bibfnamefont {V.~V.}\ \bibnamefont
  {Hovhannisyan}}, \bibinfo {author} {\bibfnamefont {J.}~\bibnamefont {Stre{\v
  c}ka}},\ and\ \bibinfo {author} {\bibfnamefont {N.~S.}\ \bibnamefont
  {Ananikian}},\ }\href {https://doi.org/10.1088/0953-8984/28/8/085401}
  {\bibfield  {journal} {\bibinfo  {journal} {Journal of Physics: Condensed
  Matter}\ }\textbf {\bibinfo {volume} {28}},\ \bibinfo {pages} {085401}
  (\bibinfo {year} {2016})}\BibitemShut {NoStop}%
\bibitem [{\citenamefont {Carvalho}\ \emph {et~al.}(2018)\citenamefont
  {Carvalho}, \citenamefont {Torrico}, \citenamefont {De~Souza}, \citenamefont
  {Rojas},\ and\ \citenamefont {Rojas}}]{Carvalho2018JMMM}%
  \BibitemOpen
  \bibfield  {author} {\bibinfo {author} {\bibfnamefont {I.}~\bibnamefont
  {Carvalho}}, \bibinfo {author} {\bibfnamefont {J.}~\bibnamefont {Torrico}},
  \bibinfo {author} {\bibfnamefont {S.}~\bibnamefont {De~Souza}}, \bibinfo
  {author} {\bibfnamefont {M.}~\bibnamefont {Rojas}},\ and\ \bibinfo {author}
  {\bibfnamefont {O.}~\bibnamefont {Rojas}},\ }\href
  {https://doi.org/10.1016/j.jmmm.2018.06.018} {\bibfield  {journal} {\bibinfo
  {journal} {Journal of Magnetism and Magnetic Materials}\ }\textbf {\bibinfo
  {volume} {465}},\ \bibinfo {pages} {323} (\bibinfo {year}
  {2018})}\BibitemShut {NoStop}%
\bibitem [{\citenamefont {Carvalho}\ \emph {et~al.}(2019)\citenamefont
  {Carvalho}, \citenamefont {Torrico}, \citenamefont {De~Souza}, \citenamefont
  {Rojas},\ and\ \citenamefont {Derzhko}}]{Carvalho2019AP}%
  \BibitemOpen
  \bibfield  {author} {\bibinfo {author} {\bibfnamefont {I.}~\bibnamefont
  {Carvalho}}, \bibinfo {author} {\bibfnamefont {J.}~\bibnamefont {Torrico}},
  \bibinfo {author} {\bibfnamefont {S.}~\bibnamefont {De~Souza}}, \bibinfo
  {author} {\bibfnamefont {O.}~\bibnamefont {Rojas}},\ and\ \bibinfo {author}
  {\bibfnamefont {O.}~\bibnamefont {Derzhko}},\ }\href
  {https://doi.org/10.1016/j.aop.2019.01.001} {\bibfield  {journal} {\bibinfo
  {journal} {Annals of Physics}\ }\textbf {\bibinfo {volume} {402}},\ \bibinfo
  {pages} {45} (\bibinfo {year} {2019})}\BibitemShut {NoStop}%
\bibitem [{\citenamefont {Rojas}\ \emph {et~al.}(2016)\citenamefont {Rojas},
  \citenamefont {Stre{\v c}ka},\ and\ \citenamefont {De~Souza}}]{Rojas2016SSC}%
  \BibitemOpen
  \bibfield  {author} {\bibinfo {author} {\bibfnamefont {O.}~\bibnamefont
  {Rojas}}, \bibinfo {author} {\bibfnamefont {J.}~\bibnamefont {Stre{\v
  c}ka}},\ and\ \bibinfo {author} {\bibfnamefont {S.}~\bibnamefont
  {De~Souza}},\ }\href {https://doi.org/10.1016/j.ssc.2016.08.002} {\bibfield
  {journal} {\bibinfo  {journal} {Solid State Communications}\ }\textbf
  {\bibinfo {volume} {246}},\ \bibinfo {pages} {68} (\bibinfo {year}
  {2016})}\BibitemShut {NoStop}%
\bibitem [{\citenamefont {Hutak}\ \emph {et~al.}(2021)\citenamefont {Hutak},
  \citenamefont {Krokhmalskii}, \citenamefont {Rojas}, \citenamefont {Martins
  De~Souza},\ and\ \citenamefont {Derzhko}}]{Hutak2021PLA}%
  \BibitemOpen
  \bibfield  {author} {\bibinfo {author} {\bibfnamefont {T.}~\bibnamefont
  {Hutak}}, \bibinfo {author} {\bibfnamefont {T.}~\bibnamefont {Krokhmalskii}},
  \bibinfo {author} {\bibfnamefont {O.}~\bibnamefont {Rojas}}, \bibinfo
  {author} {\bibfnamefont {S.}~\bibnamefont {Martins De~Souza}},\ and\ \bibinfo
  {author} {\bibfnamefont {O.}~\bibnamefont {Derzhko}},\ }\href
  {https://doi.org/10.1016/j.physleta.2020.127020} {\bibfield  {journal}
  {\bibinfo  {journal} {Physics Letters A}\ }\textbf {\bibinfo {volume}
  {387}},\ \bibinfo {pages} {127020} (\bibinfo {year} {2021})}\BibitemShut
  {NoStop}%
\bibitem [{\citenamefont {Stre{\v c}ka}\ \emph {et~al.}(2016)\citenamefont
  {Stre{\v c}ka}, \citenamefont {Al{\'e}cio}, \citenamefont {Lyra},\ and\
  \citenamefont {Rojas}}]{Strecka2016JMMM}%
  \BibitemOpen
  \bibfield  {author} {\bibinfo {author} {\bibfnamefont {J.}~\bibnamefont
  {Stre{\v c}ka}}, \bibinfo {author} {\bibfnamefont {R.~C.}\ \bibnamefont
  {Al{\'e}cio}}, \bibinfo {author} {\bibfnamefont {M.~L.}\ \bibnamefont
  {Lyra}},\ and\ \bibinfo {author} {\bibfnamefont {O.}~\bibnamefont {Rojas}},\
  }\href {https://doi.org/10.1016/j.jmmm.2016.02.095} {\bibfield  {journal}
  {\bibinfo  {journal} {Journal of Magnetism and Magnetic Materials}\ }\textbf
  {\bibinfo {volume} {409}},\ \bibinfo {pages} {124} (\bibinfo {year}
  {2016})}\BibitemShut {NoStop}%
\bibitem [{\citenamefont {G{\'a}lisov{\'a}}\ and\ \citenamefont
  {Stre{\v{c}}ka}(2015)}]{galisova2015vigorous}%
  \BibitemOpen
  \bibfield  {author} {\bibinfo {author} {\bibfnamefont {L.}~\bibnamefont
  {G{\'a}lisov{\'a}}}\ and\ \bibinfo {author} {\bibfnamefont {J.}~\bibnamefont
  {Stre{\v{c}}ka}},\ }\href {https://doi.org/10.1103/PhysRevE.91.022134}
  {\bibfield  {journal} {\bibinfo  {journal} {Physical Review E}\ }\textbf
  {\bibinfo {volume} {91}},\ \bibinfo {pages} {022134} (\bibinfo {year}
  {2015})}\BibitemShut {NoStop}%
\bibitem [{\citenamefont {Rojas}\ \emph {et~al.}(2020)\citenamefont {Rojas},
  \citenamefont {Stre{\v c}ka}, \citenamefont {Derzhko},\ and\ \citenamefont
  {De~Souza}}]{Rojas2020JPCM}%
  \BibitemOpen
  \bibfield  {author} {\bibinfo {author} {\bibfnamefont {O.}~\bibnamefont
  {Rojas}}, \bibinfo {author} {\bibfnamefont {J.}~\bibnamefont {Stre{\v c}ka}},
  \bibinfo {author} {\bibfnamefont {O.}~\bibnamefont {Derzhko}},\ and\ \bibinfo
  {author} {\bibfnamefont {S.~M.}\ \bibnamefont {De~Souza}},\ }\href
  {https://doi.org/10.1088/1361-648X/ab4acc} {\bibfield  {journal} {\bibinfo
  {journal} {Journal of Physics: Condensed Matter}\ }\textbf {\bibinfo {volume}
  {32}},\ \bibinfo {pages} {035804} (\bibinfo {year} {2020})}\BibitemShut
  {NoStop}%
\bibitem [{\citenamefont {Pimenta}\ \emph {et~al.}(2022)\citenamefont
  {Pimenta}, \citenamefont {Rojas},\ and\ \citenamefont
  {De~Souza}}]{Pimenta2022JMMM}%
  \BibitemOpen
  \bibfield  {author} {\bibinfo {author} {\bibfnamefont {R.}~\bibnamefont
  {Pimenta}}, \bibinfo {author} {\bibfnamefont {O.}~\bibnamefont {Rojas}},\
  and\ \bibinfo {author} {\bibfnamefont {S.}~\bibnamefont {De~Souza}},\ }\href
  {https://doi.org/10.1016/j.jmmm.2022.169070} {\bibfield  {journal} {\bibinfo
  {journal} {Journal of Magnetism and Magnetic Materials}\ }\textbf {\bibinfo
  {volume} {550}},\ \bibinfo {pages} {169070} (\bibinfo {year}
  {2022})}\BibitemShut {NoStop}%
\bibitem [{\citenamefont {Panov}\ and\ \citenamefont
  {Rojas}(2023)}]{Panov2023PRE}%
  \BibitemOpen
  \bibfield  {author} {\bibinfo {author} {\bibfnamefont {Y.}~\bibnamefont
  {Panov}}\ and\ \bibinfo {author} {\bibfnamefont {O.}~\bibnamefont {Rojas}},\
  }\href {https://doi.org/10.1103/PhysRevE.108.044144} {\bibfield  {journal}
  {\bibinfo  {journal} {Physical Review E}\ }\textbf {\bibinfo {volume}
  {108}},\ \bibinfo {pages} {044144} (\bibinfo {year} {2023})}\BibitemShut
  {NoStop}%
\bibitem [{\citenamefont {Coulon}\ \emph {et~al.}(2006)\citenamefont {Coulon},
  \citenamefont {Miyasaka},\ and\ \citenamefont {Cl{\'{e}}rac}}]{Coulon2006}%
  \BibitemOpen
  \bibfield  {author} {\bibinfo {author} {\bibfnamefont {C.}~\bibnamefont
  {Coulon}}, \bibinfo {author} {\bibfnamefont {H.}~\bibnamefont {Miyasaka}},\
  and\ \bibinfo {author} {\bibfnamefont {R.}~\bibnamefont {Cl{\'{e}}rac}},\
  }in\ \href {https://doi.org/10.1007/430_030} {\emph {\bibinfo {booktitle}
  {Structure and Bonding}}},\ Vol.\ \bibinfo {volume} {122}\ (\bibinfo
  {publisher} {Springer Berlin Heidelberg},\ \bibinfo {address} {Berlin,
  Heidelberg},\ \bibinfo {year} {2006})\ pp.\ \bibinfo {pages}
  {163--206}\BibitemShut {NoStop}%
\bibitem [{\citenamefont {Zhang}\ \emph {et~al.}(2013)\citenamefont {Zhang},
  \citenamefont {Ishikawa}, \citenamefont {Breedlove},\ and\ \citenamefont
  {Yamashita}}]{Zhang2013}%
  \BibitemOpen
  \bibfield  {author} {\bibinfo {author} {\bibfnamefont {W.-X.}\ \bibnamefont
  {Zhang}}, \bibinfo {author} {\bibfnamefont {R.}~\bibnamefont {Ishikawa}},
  \bibinfo {author} {\bibfnamefont {B.}~\bibnamefont {Breedlove}},\ and\
  \bibinfo {author} {\bibfnamefont {M.}~\bibnamefont {Yamashita}},\ }\href
  {https://doi.org/10.1039/c2ra22675h} {\bibfield  {journal} {\bibinfo
  {journal} {RSC Advances}\ }\textbf {\bibinfo {volume} {3}},\ \bibinfo {pages}
  {3772} (\bibinfo {year} {2013})}\BibitemShut {NoStop}%
\bibitem [{\citenamefont {Liu}\ \emph {et~al.}(2013)\citenamefont {Liu},
  \citenamefont {Zheng}, \citenamefont {Kang}, \citenamefont {Shiota},
  \citenamefont {Hayami}, \citenamefont {Mito}, \citenamefont {Sato},
  \citenamefont {Yoshizawa}, \citenamefont {Kanegawa},\ and\ \citenamefont
  {Duan}}]{Liu2013}%
  \BibitemOpen
  \bibfield  {author} {\bibinfo {author} {\bibfnamefont {T.}~\bibnamefont
  {Liu}}, \bibinfo {author} {\bibfnamefont {H.}~\bibnamefont {Zheng}}, \bibinfo
  {author} {\bibfnamefont {S.}~\bibnamefont {Kang}}, \bibinfo {author}
  {\bibfnamefont {Y.}~\bibnamefont {Shiota}}, \bibinfo {author} {\bibfnamefont
  {S.}~\bibnamefont {Hayami}}, \bibinfo {author} {\bibfnamefont
  {M.}~\bibnamefont {Mito}}, \bibinfo {author} {\bibfnamefont {O.}~\bibnamefont
  {Sato}}, \bibinfo {author} {\bibfnamefont {K.}~\bibnamefont {Yoshizawa}},
  \bibinfo {author} {\bibfnamefont {S.}~\bibnamefont {Kanegawa}},\ and\
  \bibinfo {author} {\bibfnamefont {C.}~\bibnamefont {Duan}},\ }\href
  {https://doi.org/10.1038/ncomms3826} {\bibfield  {journal} {\bibinfo
  {journal} {Nature Communications}\ }\textbf {\bibinfo {volume} {4}},\
  \bibinfo {pages} {2826} (\bibinfo {year} {2013})}\BibitemShut {NoStop}%
\bibitem [{\citenamefont {Moskvin}(2023)}]{Moskvin2023}%
  \BibitemOpen
  \bibfield  {author} {\bibinfo {author} {\bibfnamefont {A.~S.}\ \bibnamefont
  {Moskvin}},\ }\href {https://doi.org/10.3390/magnetochemistry9110224}
  {\bibfield  {journal} {\bibinfo  {journal} {Magnetochemistry}\ }\textbf
  {\bibinfo {volume} {9}},\ \bibinfo {pages} {1} (\bibinfo {year} {2023})},\
  \Eprint {https://arxiv.org/abs/2306.06612} {2306.06612} \BibitemShut
  {NoStop}%
\bibitem [{\citenamefont {Panov}(2022)}]{Panov2022PRE}%
  \BibitemOpen
  \bibfield  {author} {\bibinfo {author} {\bibfnamefont {Y.}~\bibnamefont
  {Panov}},\ }\href {https://doi.org/10.1103/PhysRevE.106.054111} {\bibfield
  {journal} {\bibinfo  {journal} {Physical Review E}\ }\textbf {\bibinfo
  {volume} {106}},\ \bibinfo {pages} {054111} (\bibinfo {year}
  {2022})}\BibitemShut {NoStop}%
\bibitem [{\citenamefont {Panov}\ \emph
  {et~al.}(2017{\natexlab{a}})\citenamefont {Panov}, \citenamefont {Moskvin},
  \citenamefont {Chikov},\ and\ \citenamefont {Budrin}}]{Panov2017JLTP}%
  \BibitemOpen
  \bibfield  {author} {\bibinfo {author} {\bibfnamefont {Y.~D.}\ \bibnamefont
  {Panov}}, \bibinfo {author} {\bibfnamefont {A.~S.}\ \bibnamefont {Moskvin}},
  \bibinfo {author} {\bibfnamefont {A.~A.}\ \bibnamefont {Chikov}},\ and\
  \bibinfo {author} {\bibfnamefont {K.~S.}\ \bibnamefont {Budrin}},\ }\href
  {https://doi.org/10.1007/s10909-017-1743-9} {\bibfield  {journal} {\bibinfo
  {journal} {Journal of Low Temperature Physics}\ }\textbf {\bibinfo {volume}
  {187}},\ \bibinfo {pages} {646} (\bibinfo {year}
  {2017}{\natexlab{a}})}\BibitemShut {NoStop}%
\bibitem [{\citenamefont {Panov}\ \emph
  {et~al.}(2017{\natexlab{b}})\citenamefont {Panov}, \citenamefont {Budrin},
  \citenamefont {Chikov},\ and\ \citenamefont {Moskvin}}]{Panov2017JETPlett}%
  \BibitemOpen
  \bibfield  {author} {\bibinfo {author} {\bibfnamefont {Y.~D.}\ \bibnamefont
  {Panov}}, \bibinfo {author} {\bibfnamefont {K.~S.}\ \bibnamefont {Budrin}},
  \bibinfo {author} {\bibfnamefont {A.~A.}\ \bibnamefont {Chikov}},\ and\
  \bibinfo {author} {\bibfnamefont {A.~S.}\ \bibnamefont {Moskvin}},\ }\href
  {https://doi.org/10.1134/S002136401719002X} {\bibfield  {journal} {\bibinfo
  {journal} {JETP Letters}\ }\textbf {\bibinfo {volume} {106}},\ \bibinfo
  {pages} {440} (\bibinfo {year} {2017}{\natexlab{b}})}\BibitemShut {NoStop}%
\bibitem [{\citenamefont {Panov}\ \emph
  {et~al.}(2019{\natexlab{a}})\citenamefont {Panov}, \citenamefont {Ulitko},
  \citenamefont {Budrin}, \citenamefont {Chikov},\ and\ \citenamefont
  {Moskvin}}]{Panov2019JMMM}%
  \BibitemOpen
  \bibfield  {author} {\bibinfo {author} {\bibfnamefont {Y.}~\bibnamefont
  {Panov}}, \bibinfo {author} {\bibfnamefont {V.}~\bibnamefont {Ulitko}},
  \bibinfo {author} {\bibfnamefont {K.}~\bibnamefont {Budrin}}, \bibinfo
  {author} {\bibfnamefont {A.}~\bibnamefont {Chikov}},\ and\ \bibinfo {author}
  {\bibfnamefont {A.}~\bibnamefont {Moskvin}},\ }\href
  {https://doi.org/10.1016/j.jmmm.2019.01.049} {\bibfield  {journal} {\bibinfo
  {journal} {Journal of Magnetism and Magnetic Materials}\ }\textbf {\bibinfo
  {volume} {477}},\ \bibinfo {pages} {162} (\bibinfo {year}
  {2019}{\natexlab{a}})}\BibitemShut {NoStop}%
\bibitem [{\citenamefont {Panov}\ \emph
  {et~al.}(2019{\natexlab{b}})\citenamefont {Panov}, \citenamefont {Budrin},
  \citenamefont {Ulitko}, \citenamefont {Chikov},\ and\ \citenamefont
  {Moskvin}}]{Panov2019JSNM}%
  \BibitemOpen
  \bibfield  {author} {\bibinfo {author} {\bibfnamefont {Y.~D.}\ \bibnamefont
  {Panov}}, \bibinfo {author} {\bibfnamefont {K.~S.}\ \bibnamefont {Budrin}},
  \bibinfo {author} {\bibfnamefont {V.~A.}\ \bibnamefont {Ulitko}}, \bibinfo
  {author} {\bibfnamefont {A.~A.}\ \bibnamefont {Chikov}},\ and\ \bibinfo
  {author} {\bibfnamefont {A.~S.}\ \bibnamefont {Moskvin}},\ }\href
  {https://doi.org/10.1007/s10948-018-4892-4} {\bibfield  {journal} {\bibinfo
  {journal} {Journal of Superconductivity and Novel Magnetism}\ }\textbf
  {\bibinfo {volume} {32}},\ \bibinfo {pages} {1831} (\bibinfo {year}
  {2019}{\natexlab{b}})}\BibitemShut {NoStop}%
\bibitem [{\citenamefont {Yasinskaya}\ \emph
  {et~al.}(2020{\natexlab{a}})\citenamefont {Yasinskaya}, \citenamefont
  {Ulitko},\ and\ \citenamefont {Panov}}]{Yasinskaya2020PSS}%
  \BibitemOpen
  \bibfield  {author} {\bibinfo {author} {\bibfnamefont {D.~N.}\ \bibnamefont
  {Yasinskaya}}, \bibinfo {author} {\bibfnamefont {V.~A.}\ \bibnamefont
  {Ulitko}},\ and\ \bibinfo {author} {\bibfnamefont {Y.~D.}\ \bibnamefont
  {Panov}},\ }\href {https://doi.org/10.1134/S1063783420090346} {\bibfield
  {journal} {\bibinfo  {journal} {Physics of the Solid State}\ }\textbf
  {\bibinfo {volume} {62}},\ \bibinfo {pages} {1713} (\bibinfo {year}
  {2020}{\natexlab{a}})}\BibitemShut {NoStop}%
\bibitem [{\citenamefont {Yasinskaya}\ \emph
  {et~al.}(2020{\natexlab{b}})\citenamefont {Yasinskaya}, \citenamefont
  {Ulitko}, \citenamefont {Chikov},\ and\ \citenamefont
  {Panov}}]{Yasinskaya2020APPA}%
  \BibitemOpen
  \bibfield  {author} {\bibinfo {author} {\bibfnamefont {D.}~\bibnamefont
  {Yasinskaya}}, \bibinfo {author} {\bibfnamefont {V.}~\bibnamefont {Ulitko}},
  \bibinfo {author} {\bibfnamefont {A.}~\bibnamefont {Chikov}},\ and\ \bibinfo
  {author} {\bibfnamefont {Y.}~\bibnamefont {Panov}},\ }\href
  {https://doi.org/10.12693/APhysPolA.137.979} {\bibfield  {journal} {\bibinfo
  {journal} {Acta Physica Polonica A}\ }\textbf {\bibinfo {volume} {137}},\
  \bibinfo {pages} {979} (\bibinfo {year} {2020}{\natexlab{b}})}\BibitemShut
  {NoStop}%
\bibitem [{\citenamefont {Yasinskaya}\ \emph {et~al.}(2021)\citenamefont
  {Yasinskaya}, \citenamefont {Ulitko},\ and\ \citenamefont
  {Panov}}]{Yasinskaya2021PSS}%
  \BibitemOpen
  \bibfield  {author} {\bibinfo {author} {\bibfnamefont {D.~N.}\ \bibnamefont
  {Yasinskaya}}, \bibinfo {author} {\bibfnamefont {V.~A.}\ \bibnamefont
  {Ulitko}},\ and\ \bibinfo {author} {\bibfnamefont {Y.~D.}\ \bibnamefont
  {Panov}},\ }\href {https://doi.org/10.1134/S1063783421090468} {\bibfield
  {journal} {\bibinfo  {journal} {Physics of the Solid State}\ }\textbf
  {\bibinfo {volume} {63}},\ \bibinfo {pages} {1588} (\bibinfo {year}
  {2021})}\BibitemShut {NoStop}%
\bibitem [{\citenamefont {Yasinskaya}\ \emph {et~al.}(2022)\citenamefont
  {Yasinskaya}, \citenamefont {Ulitko},\ and\ \citenamefont
  {Panov}}]{Yasinskaya2022IEEE}%
  \BibitemOpen
  \bibfield  {author} {\bibinfo {author} {\bibfnamefont {D.~N.}\ \bibnamefont
  {Yasinskaya}}, \bibinfo {author} {\bibfnamefont {V.~A.}\ \bibnamefont
  {Ulitko}},\ and\ \bibinfo {author} {\bibfnamefont {Y.~D.}\ \bibnamefont
  {Panov}},\ }\href {https://doi.org/10.1109/TMAG.2021.3082178} {\bibfield
  {journal} {\bibinfo  {journal} {IEEE Transactions on Magnetics}\ }\textbf
  {\bibinfo {volume} {58}},\ \bibinfo {pages} {1} (\bibinfo {year}
  {2022})}\BibitemShut {NoStop}%
\bibitem [{\citenamefont {Moskvin}(2011)}]{Moskvin2011PRB}%
  \BibitemOpen
  \bibfield  {author} {\bibinfo {author} {\bibfnamefont {A.~S.}\ \bibnamefont
  {Moskvin}},\ }\href {https://doi.org/10.1103/PhysRevB.84.075116} {\bibfield
  {journal} {\bibinfo  {journal} {Physical Review B}\ }\textbf {\bibinfo
  {volume} {84}},\ \bibinfo {pages} {075116} (\bibinfo {year}
  {2011})}\BibitemShut {NoStop}%
\bibitem [{\citenamefont {Moskvin}(2015)}]{Moskvin2015JETP}%
  \BibitemOpen
  \bibfield  {author} {\bibinfo {author} {\bibfnamefont {A.~S.}\ \bibnamefont
  {Moskvin}},\ }\href {https://doi.org/10.1134/S1063776115090095} {\bibfield
  {journal} {\bibinfo  {journal} {Journal of Experimental and Theoretical
  Physics}\ }\textbf {\bibinfo {volume} {121}},\ \bibinfo {pages} {477}
  (\bibinfo {year} {2015})}\BibitemShut {NoStop}%
\bibitem [{\citenamefont {Panov}(2019)}]{Panov2019PMM}%
  \BibitemOpen
  \bibfield  {author} {\bibinfo {author} {\bibfnamefont {Y.~D.}\ \bibnamefont
  {Panov}},\ }\href {https://doi.org/10.1134/S0031918X19130222} {\bibfield
  {journal} {\bibinfo  {journal} {Physics of Metals and Metallography}\
  }\textbf {\bibinfo {volume} {120}},\ \bibinfo {pages} {1276} (\bibinfo {year}
  {2019})}\BibitemShut {NoStop}%
\bibitem [{\citenamefont {Moskvin}\ and\ \citenamefont
  {Panov}(2020)}]{Moskvin2020PSS}%
  \BibitemOpen
  \bibfield  {author} {\bibinfo {author} {\bibfnamefont {A.~S.}\ \bibnamefont
  {Moskvin}}\ and\ \bibinfo {author} {\bibfnamefont {Y.~D.}\ \bibnamefont
  {Panov}},\ }\href {https://doi.org/10.1134/S1063783420090206} {\bibfield
  {journal} {\bibinfo  {journal} {Physics of the Solid State}\ }\textbf
  {\bibinfo {volume} {62}},\ \bibinfo {pages} {1554} (\bibinfo {year}
  {2020})}\BibitemShut {NoStop}%
\bibitem [{\citenamefont {Moskvin}\ and\ \citenamefont
  {Panov}(2021)}]{Moskvin2021CM}%
  \BibitemOpen
  \bibfield  {author} {\bibinfo {author} {\bibfnamefont {A.}~\bibnamefont
  {Moskvin}}\ and\ \bibinfo {author} {\bibfnamefont {Y.}~\bibnamefont
  {Panov}},\ }\href {https://doi.org/10.3390/condmat6030024} {\bibfield
  {journal} {\bibinfo  {journal} {Condensed Matter}\ }\textbf {\bibinfo
  {volume} {6}},\ \bibinfo {pages} {24} (\bibinfo {year} {2021})}\BibitemShut
  {NoStop}%
\bibitem [{\citenamefont {Moskvin}\ and\ \citenamefont
  {Panov}(2022{\natexlab{a}})}]{Moskvin2022JMMM}%
  \BibitemOpen
  \bibfield  {author} {\bibinfo {author} {\bibfnamefont {A.}~\bibnamefont
  {Moskvin}}\ and\ \bibinfo {author} {\bibfnamefont {Y.}~\bibnamefont
  {Panov}},\ }\href {https://doi.org/10.1016/j.jmmm.2021.169004} {\bibfield
  {journal} {\bibinfo  {journal} {Journal of Magnetism and Magnetic Materials}\
  }\textbf {\bibinfo {volume} {550}},\ \bibinfo {pages} {169004} (\bibinfo
  {year} {2022}{\natexlab{a}})}\BibitemShut {NoStop}%
\bibitem [{\citenamefont {Moskvin}\ and\ \citenamefont
  {Panov}(2022{\natexlab{b}})}]{Moskvin2022JPCS}%
  \BibitemOpen
  \bibfield  {author} {\bibinfo {author} {\bibfnamefont {A.~S.}\ \bibnamefont
  {Moskvin}}\ and\ \bibinfo {author} {\bibfnamefont {Y.~D.}\ \bibnamefont
  {Panov}},\ }\href {https://doi.org/10.1088/1742-6596/2164/1/012014}
  {\bibfield  {journal} {\bibinfo  {journal} {Journal of Physics: Conference
  Series}\ }\textbf {\bibinfo {volume} {2164}},\ \bibinfo {pages} {012014}
  (\bibinfo {year} {2022}{\natexlab{b}})}\BibitemShut {NoStop}%
\bibitem [{\citenamefont {Stre{\v{c}}ka}\ and\ \citenamefont
  {Karl'ov{\'{a}}}(2024)}]{Strecka2024EPJB}%
  \BibitemOpen
  \bibfield  {author} {\bibinfo {author} {\bibfnamefont {J.}~\bibnamefont
  {Stre{\v{c}}ka}}\ and\ \bibinfo {author} {\bibfnamefont {K.}~\bibnamefont
  {Karl'ov{\'{a}}}},\ }\href {https://doi.org/10.1140/epjb/s10051-024-00710-7}
  {\bibfield  {journal} {\bibinfo  {journal} {The European Physical Journal B}\
  }\textbf {\bibinfo {volume} {97}},\ \bibinfo {pages} {74} (\bibinfo {year}
  {2024})}\BibitemShut {NoStop}%
\bibitem [{\citenamefont {Bounoua}\ \emph {et~al.}(2017)\citenamefont
  {Bounoua}, \citenamefont {Saint-Martin}, \citenamefont {Petit}, \citenamefont
  {Berthet}, \citenamefont {Damay}, \citenamefont {Sidis}, \citenamefont
  {Bourdarot},\ and\ \citenamefont {Pinsard-Gaudart}}]{Bounoua2017}%
  \BibitemOpen
  \bibfield  {author} {\bibinfo {author} {\bibfnamefont {D.}~\bibnamefont
  {Bounoua}}, \bibinfo {author} {\bibfnamefont {R.}~\bibnamefont
  {Saint-Martin}}, \bibinfo {author} {\bibfnamefont {S.}~\bibnamefont {Petit}},
  \bibinfo {author} {\bibfnamefont {P.}~\bibnamefont {Berthet}}, \bibinfo
  {author} {\bibfnamefont {F.}~\bibnamefont {Damay}}, \bibinfo {author}
  {\bibfnamefont {Y.}~\bibnamefont {Sidis}}, \bibinfo {author} {\bibfnamefont
  {F.}~\bibnamefont {Bourdarot}},\ and\ \bibinfo {author} {\bibfnamefont
  {L.}~\bibnamefont {Pinsard-Gaudart}},\ }\href
  {https://doi.org/10.1103/PhysRevB.95.224429} {\bibfield  {journal} {\bibinfo
  {journal} {Physical Review B}\ }\textbf {\bibinfo {volume} {95}},\ \bibinfo
  {pages} {224429} (\bibinfo {year} {2017})}\BibitemShut {NoStop}%
\bibitem [{\citenamefont {Yang}(1952)}]{Yang1952PR}%
  \BibitemOpen
  \bibfield  {author} {\bibinfo {author} {\bibfnamefont {C.~N.}\ \bibnamefont
  {Yang}},\ }\href {https://doi.org/10.1103/PhysRev.85.808} {\bibfield
  {journal} {\bibinfo  {journal} {Physical Review}\ }\textbf {\bibinfo {volume}
  {85}},\ \bibinfo {pages} {808} (\bibinfo {year} {1952})}\BibitemShut
  {NoStop}%
\bibitem [{\citenamefont {Newman}\ and\ \citenamefont
  {Barkema}(2010)}]{Newman}%
  \BibitemOpen
  \bibfield  {author} {\bibinfo {author} {\bibfnamefont {M.~E.~J.}\
  \bibnamefont {Newman}}\ and\ \bibinfo {author} {\bibfnamefont {G.~T.}\
  \bibnamefont {Barkema}},\ }\href@noop {} {\emph {\bibinfo {title} {Monte
  {Carlo} methods in statistical physics}}},\ \bibinfo {edition} {reprinted
  (with corr.)}\ ed.\ (\bibinfo  {publisher} {Clarendon Press [u.a.]},\
  \bibinfo {address} {Oxford},\ \bibinfo {year} {2010})\BibitemShut {NoStop}%
\bibitem [{\citenamefont {Batrouni}\ and\ \citenamefont
  {Scalettar}(2000)}]{Batrouni2000PRL}%
  \BibitemOpen
  \bibfield  {author} {\bibinfo {author} {\bibfnamefont {G.~G.}\ \bibnamefont
  {Batrouni}}\ and\ \bibinfo {author} {\bibfnamefont {R.~T.}\ \bibnamefont
  {Scalettar}},\ }\href {https://doi.org/10.1103/PhysRevLett.84.1599}
  {\bibfield  {journal} {\bibinfo  {journal} {Physical Review Letters}\
  }\textbf {\bibinfo {volume} {84}},\ \bibinfo {pages} {1599} (\bibinfo {year}
  {2000})}\BibitemShut {NoStop}%
\bibitem [{\citenamefont {Kapcia}\ \emph {et~al.}(2012)\citenamefont {Kapcia},
  \citenamefont {Robaszkiewicz},\ and\ \citenamefont
  {Micnas}}]{Kapcia2012JPCM}%
  \BibitemOpen
  \bibfield  {author} {\bibinfo {author} {\bibfnamefont {K.}~\bibnamefont
  {Kapcia}}, \bibinfo {author} {\bibfnamefont {S.}~\bibnamefont
  {Robaszkiewicz}},\ and\ \bibinfo {author} {\bibfnamefont {R.}~\bibnamefont
  {Micnas}},\ }\href {https://doi.org/10.1088/0953-8984/24/21/215601}
  {\bibfield  {journal} {\bibinfo  {journal} {Journal of Physics: Condensed
  Matter}\ }\textbf {\bibinfo {volume} {24}},\ \bibinfo {pages} {215601}
  (\bibinfo {year} {2012})}\BibitemShut {NoStop}%
\bibitem [{\citenamefont {Kapcia}\ and\ \citenamefont
  {Robaszkiewicz}(2013)}]{Kapcia2013JPCM}%
  \BibitemOpen
  \bibfield  {author} {\bibinfo {author} {\bibfnamefont {K.}~\bibnamefont
  {Kapcia}}\ and\ \bibinfo {author} {\bibfnamefont {S.}~\bibnamefont
  {Robaszkiewicz}},\ }\href {https://doi.org/10.1088/0953-8984/25/6/065603}
  {\bibfield  {journal} {\bibinfo  {journal} {Journal of Physics: Condensed
  Matter}\ }\textbf {\bibinfo {volume} {25}},\ \bibinfo {pages} {065603}
  (\bibinfo {year} {2013})}\BibitemShut {NoStop}%
\bibitem [{\citenamefont {Kapcia}(2013)}]{Kapcia2013JSNM}%
  \BibitemOpen
  \bibfield  {author} {\bibinfo {author} {\bibfnamefont {K.}~\bibnamefont
  {Kapcia}},\ }\href {https://doi.org/10.1007/s10948-013-2152-1} {\bibfield
  {journal} {\bibinfo  {journal} {Journal of Superconductivity and Novel
  Magnetism}\ }\textbf {\bibinfo {volume} {26}},\ \bibinfo {pages} {2647}
  (\bibinfo {year} {2013})}\BibitemShut {NoStop}%
\bibitem [{\citenamefont {Kapcia}(2015)}]{Kapcia2015JSNM}%
  \BibitemOpen
  \bibfield  {author} {\bibinfo {author} {\bibfnamefont {K.~J.}\ \bibnamefont
  {Kapcia}},\ }\href {https://doi.org/10.1007/s10948-014-2906-4} {\bibfield
  {journal} {\bibinfo  {journal} {Journal of Superconductivity and Novel
  Magnetism}\ }\textbf {\bibinfo {volume} {28}},\ \bibinfo {pages} {1289}
  (\bibinfo {year} {2015})}\BibitemShut {NoStop}%
\bibitem [{\citenamefont {Kapcia}\ \emph {et~al.}(2015)\citenamefont {Kapcia},
  \citenamefont {Murawski}, \citenamefont {K{\l }obus},\ and\ \citenamefont
  {Robaszkiewicz}}]{Kapcia2015PhysicaA}%
  \BibitemOpen
  \bibfield  {author} {\bibinfo {author} {\bibfnamefont {K.~J.}\ \bibnamefont
  {Kapcia}}, \bibinfo {author} {\bibfnamefont {S.}~\bibnamefont {Murawski}},
  \bibinfo {author} {\bibfnamefont {W.}~\bibnamefont {K{\l }obus}},\ and\
  \bibinfo {author} {\bibfnamefont {S.}~\bibnamefont {Robaszkiewicz}},\ }\href
  {https://doi.org/10.1016/j.physa.2015.05.074} {\bibfield  {journal} {\bibinfo
   {journal} {Physica A: Statistical Mechanics and its Applications}\ }\textbf
  {\bibinfo {volume} {437}},\ \bibinfo {pages} {218} (\bibinfo {year}
  {2015})}\BibitemShut {NoStop}%
\bibitem [{\citenamefont {Kapcia}\ \emph {et~al.}(2017)\citenamefont {Kapcia},
  \citenamefont {Bara{\'n}ski},\ and\ \citenamefont {Ptok}}]{Kapcia2017PRE}%
  \BibitemOpen
  \bibfield  {author} {\bibinfo {author} {\bibfnamefont {K.~J.}\ \bibnamefont
  {Kapcia}}, \bibinfo {author} {\bibfnamefont {J.}~\bibnamefont
  {Bara{\'n}ski}},\ and\ \bibinfo {author} {\bibfnamefont {A.}~\bibnamefont
  {Ptok}},\ }\href {https://doi.org/10.1103/PhysRevE.96.042104} {\bibfield
  {journal} {\bibinfo  {journal} {Physical Review E}\ }\textbf {\bibinfo
  {volume} {96}},\ \bibinfo {pages} {042104} (\bibinfo {year}
  {2017})}\BibitemShut {NoStop}%
\bibitem [{\citenamefont {Plischke}\ and\ \citenamefont
  {Bergersen}(2006)}]{Plischke2006}%
  \BibitemOpen
  \bibfield  {author} {\bibinfo {author} {\bibfnamefont {M.}~\bibnamefont
  {Plischke}}\ and\ \bibinfo {author} {\bibfnamefont {B.}~\bibnamefont
  {Bergersen}},\ }\href@noop {} {\emph {\bibinfo {title} {Equilibrium
  statistical physics}}},\ \bibinfo {edition} {3rd}\ ed.\ (\bibinfo
  {publisher} {World Scientific},\ \bibinfo {year} {2006})\BibitemShut
  {NoStop}%
\bibitem [{\citenamefont {Frobenius}(1912)}]{Frobenius1912}%
  \BibitemOpen
  \bibfield  {author} {\bibinfo {author} {\bibfnamefont {G.}~\bibnamefont
  {Frobenius}},\ }\href@noop {} {\emph {\bibinfo {title} {{\"U}ber Matrizen aus
  nicht negativen Elementen}}},\ Preussische Akademie der Wissenschaften
  Berlin: Sitzungsberichte der Preu{\ss}ischen Akademie der Wissenschaften zu
  Berlin\ (\bibinfo  {publisher} {Reichsdr.},\ \bibinfo {year}
  {1912})\BibitemShut {NoStop}%
\bibitem [{\citenamefont {Yasinskaya}\ and\ \citenamefont
  {Panov}(2024)}]{Yasinskaya2024PSS}%
  \BibitemOpen
  \bibfield  {author} {\bibinfo {author} {\bibfnamefont {D.~N.}\ \bibnamefont
  {Yasinskaya}}\ and\ \bibinfo {author} {\bibfnamefont {Y.~D.}\ \bibnamefont
  {Panov}},\ }\bibfield  {journal} {\bibinfo  {journal} {Physics of the Solid
  State}\ }\textbf {\bibinfo {volume} {66}},\ \href
  {https://doi.org/10.61011/PSS.2024.07.58978.47HH}
  {10.61011/PSS.2024.07.58978.47HH} (\bibinfo {year} {2024})\BibitemShut
  {NoStop}%
\bibitem [{\citenamefont {De~Souza}\ and\ \citenamefont
  {Rojas}(2018)}]{Souza2018SSC}%
  \BibitemOpen
  \bibfield  {author} {\bibinfo {author} {\bibfnamefont {S.}~\bibnamefont
  {De~Souza}}\ and\ \bibinfo {author} {\bibfnamefont {O.}~\bibnamefont
  {Rojas}},\ }\href {https://doi.org/10.1016/j.ssc.2017.10.006} {\bibfield
  {journal} {\bibinfo  {journal} {Solid State Communications}\ }\textbf
  {\bibinfo {volume} {269}},\ \bibinfo {pages} {131} (\bibinfo {year}
  {2018})}\BibitemShut {NoStop}%
\bibitem [{\citenamefont {Panov}(2020)}]{Panov2020JMMM}%
  \BibitemOpen
  \bibfield  {author} {\bibinfo {author} {\bibfnamefont {Y.}~\bibnamefont
  {Panov}},\ }\href {https://doi.org/10.1016/j.jmmm.2020.167224} {\bibfield
  {journal} {\bibinfo  {journal} {Journal of Magnetism and Magnetic Materials}\
  }\textbf {\bibinfo {volume} {514}},\ \bibinfo {pages} {167224} (\bibinfo
  {year} {2020})}\BibitemShut {NoStop}%
\bibitem [{\citenamefont {Korn}\ and\ \citenamefont {Korn}(2000)}]{Korn2000}%
  \BibitemOpen
  \bibfield  {author} {\bibinfo {author} {\bibfnamefont {G.~A.}\ \bibnamefont
  {Korn}}\ and\ \bibinfo {author} {\bibfnamefont {T.~M.}\ \bibnamefont
  {Korn}},\ }\href@noop {} {\emph {\bibinfo {title} {Mathematical handbook for
  scientists and engineers: definitions, theorems, and formulas for reference
  and review}}}\ (\bibinfo  {publisher} {Dover Publications},\ \bibinfo
  {address} {Mineola, New York, USA},\ \bibinfo {year} {2000})\BibitemShut
  {NoStop}%
\end{thebibliography}%

\end{document}